\documentclass[12pt]{article}
\usepackage{amsmath}
\usepackage{graphicx}
\usepackage{enumerate}
\usepackage{natbib}
\usepackage{url}

\newcommand{\blind}{1}

\newtheorem{theorem}{Theorem}
\newtheorem{corollary}{Corollary}[theorem]

\newtheorem{assumption}{Assumption}
\usepackage{multirow}
\usepackage{bm}
\usepackage{mathtools}
\DeclarePairedDelimiter\floor{\lfloor}{\rfloor}

\begin{document}

\def\spacingset#1{\renewcommand{\baselinestretch}%
{#1}\small\normalsize} \spacingset{1}


\if1\blind
{
  \title{\bf A Generalized Difference-in-Differences Estimator for Randomized Stepped-Wedge and Observational Staggered Adoption  Settings}
  \author{Lee Kennedy-Shaffer \\
    Department of Biostatistics, Yale School of Public Health}
  \maketitle
} \fi

\if0\blind
{
  \bigskip
  \bigskip
  \bigskip
  \begin{center}
    {\LARGE\bf  Generalized Difference-in-Differences Estimator for Randomized Stepped-Wedge and Observational Staggered Adoption  Settings}
\end{center}
  \medskip
} \fi

\bigskip
\begin{abstract}
Staggered treatment adoption arises in the evaluation of policy impact and implementation in many settings, including both randomized stepped-wedge trials and non-randomized quasi-experiments with panel data. In both settings, getting an interpretable, unbiased effect estimate requires careful consideration of the target estimand and possible treatment effect heterogeneities. This paper proposes a novel non-parametric approach to this estimation for either setting. By constructing an estimator using weighted averages of two-by-two difference-in-differences comparisons as building blocks, the investigator can target the desired estimand for any assumed treatment effect heterogeneities. This provides desirable bias and interpretation properties while using the comparisons efficiently to mitigate the loss of precision, without requiring correct variance specification. The methods are demonstrated for both a randomized stepped-wedge trial on the impact of novel tuberculosis diagnostic tools and an observational staggered adoption study on the effects of COVID-19 vaccine financial incentive lotteries in U.S.\ states; these are compared to analyses using previous methods. A full algorithm with R code is provided to implement this method and to compare against existing methods. The proposed method allows for high flexibility and clear targeting of desired effects, providing one solution to the bias-variance-generalizability tradeoff.
\end{abstract}

\noindent%
\vfill

\newpage
\spacingset{1.9} 
\section{Introduction}
\label{sec:intro}

Staggered treatment adoption occurs in a wide variety of settings, including both observational and randomized contexts. In observational and quasi-experimental studies, panel data methods are commonly used to analyze the effect of a policy implementation or an exogenous shock. Stepped-wedge cluster-randomized trials have also become a common approach for the analysis of health, education, or other social policies, especially those requiring a phased or gradual implementation. Across settings, the analysis of the data generated from staggered treatment adoption requires careful consideration of the desired estimand, assumptions about the treatment effect, consideration of heterogeneity across units, time periods, and treatment regimens, and appropriate consideration of variance and correlation.

This complexity has led to the development of a wide array of methods, commonly found in somewhat distinct areas: (1) the econometrics literature surrounding panel data, difference-in-differences (DID), and staggered treatment adoption; and (2) the biostatistics literature surrounding stepped-wedge trials (SWTs). Key developments in SWTs include approaches to interpret the targeted estimand (see, e.g., \cite{twisk_different_2016}), design-based considerations (see, e.g., \cite{matthews_stepped_2017} and \cite{li_planning_2024}), robust inference (see, e.g., \cite{wang_use_2017}, \cite{hughes_robust_2020}, \cite{maleyeff_assessing_2023}, and \cite{wang_how_2024}), and a variety of analytic approaches, discussed below. While development in both areas is still very much ongoing, recent reviews of the observational staggered adoption methods literature include, among others, \cite{baker_how_2022}, \cite{dechaisemartin_two-way_2023}, \cite{roth_whats_2023}, and \cite{borusyak_revisiting_2024}. Review papers in specific disciplines also highlight how to interpret the assumptions and estimands therein; e.g., in health policy, see \cite{feng_parallel_2024}, \cite{feng_difference--differences_2024}, and \cite{wang_advances_2024}.

These new developments highlight the need for careful selection of the target estimand and, for the most part, propose an identification strategy for unbiased estimation via modeling of the treatment effect in a regression framework. In the stepped-wedge setting, this is often done using a parametric or semi-parametric model for the treatment effect that can account for heterogeneity as in \cite{hooper_sample_2016}, \cite{nickless_mixed_2018}, \cite{kenny_analysis_2022}, \cite{maleyeff_assessing_2023}, \cite{lee_cluster_2024}, and \cite{wang_how_2024}. It can, however, also be done using appropriate weighting of non-parametric estimators like those proposed by \cite{thompson_robust_2018} and \cite{kennedyshaffer_novel_2020}. In addition, the stepped-wedge literature has discussed the importance of using information efficiently, by combining ``horizontal'' (within-unit) and ``vertical'' (within-time) comparisons (see, e.g., \cite{matthews_stepped_2017} and \cite{kasza_information_2019-1}).

In the observational staggered adoption setting, approaches to address the biases that can arise in two-way fixed-effects models (see, e.g., \cite{goodman-bacon_difference--differences_2021} and \cite{imai_use_2021}) have included adjusting the interpretation based on the identifying assumptions as in \cite{athey_design-based_2022}, specifying ``dynamic'' (time-varying) treatment effects as in \cite{sun_estimating_2021}, and isolating, excluding, or weighting effect estimators targeting specific group-time effects as in \cite{callaway_difference--differences_2021} and \cite{de_chaisemartin_two-way_2020}. \cite{lindner_heterogeneous_2021} highlight the relationship between these approaches, noting a similarity between the approaches of weighting specific effects in the DID and SWT settings.

This paper introduces a generalized estimator framework that can target a variety of estimands specified by investigators under different sets of assumptions about treatment effect homogeneity. This class of estimators is constructed by taking weighted averages of simple two-by-two DID estimators, and finding the minimum-variance  weighting that targets the desired estimand. The framework has the advantages of: (1) unifying methods across different settings, assumptions, and target estimands; (2) minimizing variance under certain variance-specific assumptions while preserving validity; (3) easing interpretability and improving robustness by explicitly identifying weights on the two-by-two DID estimators and on the observations themselves; and (4) permitting sensitivity analysis across heterogeneity and variance assumptions. Several existing staggered adoption estimators are special cases of this approach, allowing easy comparisons across methods.

Section~\ref{sec:meth} of this paper describes the approach to constructing the estimator and its properties, with the full algorithm in Section~\ref{sec:algo}. Section~\ref{sec:toy} provides a toy example to illuminate the algorithm in a few simple settings. Section~\ref{sec:Comps} situates several existing estimators as special cases of this framework and the assumptions they use, while noting others which do not fit within the framework. Section~\ref{sec:SWTex} re-analyzes data from a SWT on the effect of improved diagnostic methods on tuberculosis outcomes by \cite{trajman_impact_2015} using this method. Section~\ref{sec:SAex} re-analyzes data from an observational staggered adoption setting described by \cite{fuller_assessing_2022} on the effects of financial incentive lotteries to encourage uptake of COVID-19 vaccines in the United States. For both examples, comparisons are made to the results using existing methods. Section~\ref{sec:disc} comments on the advantages and limitations of the method, as well as future areas for research.

\section{Methods}
\label{sec:meth}

\subsection{Notation}

Consider a setting with $J$ periods ($j=1,\ldots,J$) and $N$ units of analysis ($i=1,\ldots,N$), which may be clusters or individual study units, for a total of $NJ$ observations. Denote by $Y_{ij}$ the outcome (or average outcome) in unit $i$ in period $j$ and by $X_{ij}$ the indicator of whether unit $i$ was treated/exposed in period $j$ (these are used interchangeably throughout, with a preference for ``treated'' for simplicity). Let $T_i = \min \{j:~X_{ij} = 1\}$ be the time period in which unit $i$ was first treated. The staggered adoption assumption is made that once treated, a unit remains treated for the duration of the study. Note that \cite{de_chaisemartin_two-way_2020} consider more general settings with identifying assumptions that do not require staggered adoption; generalizations of the following methods to those settings may be feasible but are not considered here. Note also that there may be multiple units in the same sequence, referring to a pattern of treatment (under the staggered adoption assumption, this is fully specified by $T_i$).

Without loss of generality, the unit indices are ordered from the earliest treatment adopter to the latest. That is, if $i < i'$, then $T_i \le T_{i'}$. Units with the same adoption time may be ordered in any way, as long as the ordering remains consistent.

\subsection{Two-by-two DID estimators}

For every pair of units $i \neq i'$ and pair of periods $j \neq j'$, there is a two-by-two DID estimator:
\begin{equation} \label{eqn:2x2}
D_{i,i',j,j'} = \left( Y_{ij'} - Y_{ij} \right) - \left( Y_{i' j'} - Y_{i' j} \right)
\end{equation}
Without loss of generality, consider only such estimators where $i < i'$ and $j < j'$. Note that swapping the order of either pair of indices multiplies the estimator by $-1$. There are a total of $\binom{N}{2} \binom{J}{2} = \frac{N(N-1)J(J-1)}{2}$ such estimators.

These two-by-two DID estimators can be partitioned into six mutually exclusive categories based on the treatment pattern of both units. Under the staggered treatment assumption, since $j < j'$, each cluster can either be untreated in both periods, treated in both periods, or untreated in period $j$ and treated in period $j'$. Moreover, since $i < i'$, $T_i \le T_{i'}$, so there are no periods where unit $i$ is untreated and unit $i'$ is treated. This leaves six possible combinations of treatment patterns, summarized in Table~\ref{tbl:2x2-cats}, along with the number in each category.

\begin{table}
\caption{Summary of two-by-two estimator categorization by treatment pattern.  \label{tbl:2x2-cats}}
\begin{center}
\begin{tabular}{rl|l|l}
Type & Description & Criteria & Number \\ \hline
1 & Both always-untreated & $j < j' < T_i \le T_{i'}$ & $\sum_{i=1}^{N-1} \sum_{i'=i+1}^N \binom{T_i-1}{2}$ \\
2 & Switch vs.\ always-untreated & $j < T_i \le j' < T_{i'}$ & $\sum_{i=1}^{N-1} \sum_{i'=i+1}^N (T_i-1)(T_{i'}-T_i)$ \\
3 & Always-treated vs.\ always-untreated & $T_i \le j < j' < T_{i'}$ & $\sum_{i=1}^{N-1} \sum_{i'=i+1}^N \binom{T_{i'}- T_i}{2}$ \\
4 & Both switch & $j < T_i \le T_{i'} \le j'$ & $\sum_{i=1}^{N-1} \sum_{i'=i+1}^N (T_i - 1) \left(J-(T_{i'}-1) \right)$ \\
5 & Always-treated vs.\ switch & $T_i \le j < T_{i'} \le j'$ & $\sum_{i=1}^{N-1} \sum_{i'=i+1}^N (T_{i'} - T_i) \left( J-(T_{i'}-1 ) \right)$ \\
6 & Both always-treated & $T_i \le T_{i'} \le j < j'$ & $\sum_{i=1}^{N-1} \sum_{i'=i+1}^N \binom{J-(T_{i'}-1)}{2}$ \\
\end{tabular}
\end{center}
\end{table}

\subsection{Expected value of two-by-two DID estimators}

This paper focuses on additive treatment effects, so the effect of treatment in period $j$ on unit $i$, which first adopted treatment in period $T_i$, denoted $\theta_{i,T_i,j}$, is given by:
\begin{equation} \label{eqn:theta}
\theta_{i,T_i,j} = E[ Y_{ij}(1) ] - E[Y_{ij}(0)],
\end{equation}
where $Y_{ij}(1)$ is the potential outcome for unit $i$ in period $j$ if the unit is first treated in period $T_i \le j$ and $Y_{ij}(0)$ is the potential outcome for unit $i$ in period $j$ if the unit is first treated after period $j$ or never treated. Note that multiplicative treatment effects can be considered by using a log transformation on the outcome. If the assumptions hold on the multiplicative scale, the estimator can then be computed using the log-transformed outcomes as the $Y_{ij}$ values. This has been discussed in staggered adoption settings (see, e.g., \cite{kahn-lang_promise_2020} and \cite{goodman-bacon_using_2020}) and in SWT settings (see, e.g., \cite{kennedyshaffer_novel_2020} and \cite{kennedy-shaffer_statistical_2020}).

To ensure unbiased estimation, the following assumptions (no spillover and no anticipation, respectively) are also needed; versions of these are common in both the staggered adoption and SWT literatures.
\begin{assumption} \label{as:no-spill}
For all $i,j$, $Y_{ij}(1)$ and $Y_{ij}(0)$ are independent of $X_{i'j'}$ for any $i' \neq i$ and any $j'$.
\end{assumption}
\begin{assumption} \label{as:no-ant}
For all $j < T_i$, $Y_{ij} = Y_{ij}(0)$.
\end{assumption}

From the definitions, then, the expectation of each two-by-two DID estimator can be found as follows.
\begin{theorem} \label{thm:expn-did}
Under Assumptions \ref{as:no-spill} and \ref{as:no-ant}, for all $i,i',j,j'$:
\begin{align*} E \left[ D_{i,i',j,j'} \right] &= \left(E[Y_{ij'}(0)] - E[Y_{ij}(0)] \right) - \left(E[Y_{i'j'}(0)] - E[Y_{i'j}(0)] \right) \\
&\qquad + \left( \theta_{i,T_i,j'} X_{ij'} - \theta_{i,T_i,j} X_{ij} \right) - \left( \theta_{i',T_{i'},j'} X_{i'j'} - \theta_{i',T_{i'},j} X_{i'j} \right). \end{align*}
\end{theorem}

This expectation can be simplified under a further parallel trends assumption.
\begin{assumption} \label{as:par-trends}
For any $i \neq i'$ and $j < j'$, $E[Y_{ij'}(0)] - E[Y_{ij}(0)]  = E[Y_{i'j'}(0)] - E[Y_{i'j}(0)]$.
\end{assumption}
Note that this aligns with one version of the parallel (or common) trends assumption used in the staggered adoption literature (see, e.g., \cite{cunningham_causal_2021}, p. 435), but other forms and statements of it exist as well. This can be implied by randomization of the treatment adoption sequences, as in the SWT case, as that would imply exchangeability of outcomes which necessarily implies parallel trends.
\begin{corollary} \label{corr:expn-did-par}
Under Assumptions~\ref{as:no-spill}--\ref{as:par-trends}, for all $i,i',j,j'$:
\[ E \left[ D_{i,i',j,j'} \right] = \left( \theta_{i,T_i,j'} X_{ij'} - \theta_{i,T_i,j} X_{ij} \right) - \left( \theta_{i',T_{i'},j'} X_{i'j'} - \theta_{i',T_{i'},j} X_{i'j} \right). \]
\end{corollary}

Further simplification of this expectation depends on assumptions about the heterogeneity of the treatment effects by unit, time-on-treatment (i.e., exposure time), and period (i.e., calendar time). Many assumption settings are feasible; this paper considers five and defines a simplified treatment effect notation for each; these are described in Table~\ref{tbl:assnset}.

\begin{table}[!ht]
\caption{Treatment effect heterogeneity assumption settings.\label{tbl:assnset}}
\begin{center}
\begin{tabular}{r|p{40mm}p{45mm}cc|}
 & Assumed\newline homogeneities & Allowed heterogeneities & Notation & Effect values \\
 \hline
 S1 & None & Unit ($i$)\newline Exposure time ($a$)\newline Calendar time ($j$) & $\theta^{(1)}_{i,j,a}$ & $\theta_{i,T_i,j} = \theta^{(1)}_{i,j,j-T_i+1}$ \\
 S2 & Unit & Exposure time ($a$)\newline Calendar time ($j$) & $\theta^{(2)}_{j,a}$ & $\theta_{i,T_i,j} = \theta^{(2)}_{j,j-T_i+1}$ \\
 S3 & Unit, calendar time & Exposure time ($a$) & $\theta^{(3)}_{a}$ & $\theta_{i,T_i,j} = \theta^{(3)}_{j-T_i+1}$ \\
 S4 & Unit, exposure time & Calendar time ($j$) & $\theta^{(4)}_{j}$ & $\theta_{i,T_i,j} = \theta^{(4)}_{j}$ \\
 S5 & Unit, exposure time,\newline calendar time & None & $\theta^{(5)}$ & $\theta_{i,T_i,j} = \theta^{(5)}$ \\
 \hline
\end{tabular}
\end{center}
\end{table}

Under these assumptions, the expected value given by Corollary~\ref{corr:expn-did-par} can be simplified for each type of two-by-two DID estimator. The results are given in Table~\ref{tbl:exp2}.
\begin{table}
\caption{Expected values of two-by-two estimators by treatment pattern category and treatment effect heterogeneity assumption setting.  \label{tbl:exp2}}
\begin{center}
\begin{tabular}{r|l|l|l|l|l|}
Type & Assumption S1 & S2 & S3 & S4 & S5 \\ \hline
1 & $0$ & $0$ & $0$ & $0$ & $0$ \\
2 & $\theta^{(1)}_{i,T_i,j'}$ & $\theta^{(2)}_{j',j' - (T_i - 1)}$ & $\theta^{(3)}_{j' - (T_i - 1)}$ & $\theta^{(4)}_{j'}$ & $\theta^{(5)}$ \\
3 & $\theta^{(1)}_{i,T_i,j'} - \theta^{(1)}_{i,T_i,j}$ & $\theta^{(2)}_{j',j'-(T_i-1)} - \theta^{(2)}_{j,j - (T_i-1)}$ & $\theta^{(3)}_{j'-(T_i-1)} - \theta^{(3)}_{j - (T_i-1)}$ & $\theta^{(4)}_{j'} - \theta^{(4)}_j$ & $0$ \\
4 & $\theta^{(1)}_{i,T_i,j'} - \theta^{(1)}_{i',T_{i'},j'}$ & $\theta^{(2)}_{j' - (T_i - 1)} - \theta^{(2)}_{j' - (T_{i'} - 1)}$ & $\theta^{(3)}_{j' - (T_i - 1)} - \theta^{(3)}_{j' - (T_{i'} - 1)}$ & $0$ & $0$  \\
\multirow{3}{*}{5} & \multirow{3}{3.1cm}{$\theta^{(1)}_{i,T_i,j'} - \theta^{(1)}_{i,T_i,j} - \left( \theta^{(1)}_{i',T_{i'},j'} \right)$} & \multirow{3}{4.6cm}{$\theta^{(2)}_{j',j' - (T_i - 1)} - \theta^{(2)}_{j,j - (T_i - 1)} - \left(\theta^{(2)}_{j',j' - (T_{i'} - 1)} \right)$} & \multirow{3}{4.1cm}{$\theta^{(3)}_{j' - (T_i - 1)} - \theta^{(3)}_{j - (T_i - 1)} - \left( \theta^{(3)}_{j' - (T_{i'} - 1)} \right)$} & \multirow{3}{*}{$-\theta^{(4)}_j$} & \multirow{3}{*}{$-\theta^{(5)}$} \\
&&&&& \\
&&&&& \\
\multirow{3}{*}{6} & \multirow{3}{3.1cm}{$\theta^{(1)}_{i,T_i,j'} - \theta^{(1)}_{i,T_i,j} - \left( \theta^{(1)}_{i',T_{i'},j'} - \theta^{(1)}_{i',T_{i'},j} \right)$} & \multirow{3}{4.6cm}{$\theta^{(2)}_{j',j' - (T_i - 1)} - \theta^{(2)}_{j,j - (T_i - 1)} - \left( \theta^{(2)}_{j',j' - (T_{i'} - 1)} - \theta^{(2)}_{j,j - (T_{i'} - 1)} \right)$} & \multirow{3}{4.1cm}{$\theta^{(3)}_{j' - (T_i - 1)} - \theta^{(3)}_{j - (T_i - 1)} - \left( \theta^{(3)}_{j' - (T_{i'} - 1)} - \theta^{(3)}_{j - (T_{i'} - 1)} \right)$} & \multirow{3}{*}{$0$} & \multirow{3}{*}{$0$} \\
&&&&& \\
&&&&& \\
\end{tabular}
\end{center}
\end{table}

\subsection{Overall estimator form} \label{sec:meth-OEF}
I propose a class of estimators constructed by a weighted sum of the two-by-two DID estimators $D_{i,i',j,j'}$ as follows:
\begin{equation} \label{eqn:theta-hat}
\hat{\theta} = \sum_{i=1}^{N-1} \sum_{i'=i+1}^N \sum_{j=1}^{J-1} \sum_{j'=j+1}^N w_{i,i',j,j'} D_{i,i',j,j'},
\end{equation}
with no general restrictions on the weights $w_{i,i',j,j'}$. Letting:
\[ \bm{d} = \begin{pmatrix} D_{1,2,1,2} & D_{1,2,1,3} & \cdots & D_{1,2,1,J} & D_{1,3,1,2} & \cdots & D_{N-1,N,J-1,J} \end{pmatrix}^T \]
be the $\binom{N}{2} \binom{J}{2}$-vector of two-by-two DID estimators ordered by unit $i$, unit $i'$, period $j$, period $j'$, respectively, and $\bm{w} = \begin{pmatrix} w_{1,2,1,2} & \cdots & w_{N-1,N,J-1,J} \end{pmatrix}^T$ be the corresponding vector of weights, the overall estimator can be written as:
\begin{equation} \label{eqn:theta-hat-v}
\hat{\theta} = \bm{w}^T \bm{d}
\end{equation}

Further, define $\bm{y} = \begin{pmatrix} Y_{1,1} & Y_{1,2} & \cdots & Y_{1,J} & Y_{2,1} & \cdots & Y_{N,J} \end{pmatrix}^T$ as the $NJ$-vector of observed outcomes ordered by unit $i$ and period $j$, respectively. Finally, let $\bm{A}$ be the $\binom{N}{2} \binom{J}{2} \times NJ$ matrix such that $\bm{d} = \bm{A} \bm{y}$. Note that each row of $\bm{A}$ corresponds to a unique two-by-two DID estimator and includes two entries of 1 and two entries of -1, with the remaining entries all 0. An algorithm to generate $\bm{A}$ for any $N$ and $J$ is given in Appendix~\ref{appx:A_const}.

If the weight vector $\bm{w}$ is independent of the outcomes $\bm{y}$, then:
\begin{equation} \label{eqn:e-theta-hat}
E[\hat{\theta}] = \bm{w}^T E[\bm{d}] = \sum_{i=1}^{N-1} \sum_{i'=i+1}^N \sum_{j=1}^{J-1} \sum_{j'=j+1}^N w_{i,i',j,j'} E[D_{i,i',j,j'}].
\end{equation}
This can be simplified under any assumption setting using the results shown in Table~\ref{tbl:exp2}.

\subsection{Unbiased estimation of target estimand}
For any assumption setting, denote the vector of all unique treatment effects by $\bm{\theta}$, with a superscript to clarify the assumption setting if desired. For concreteness, order the treatment effects by cluster $i$, period $j$, and exposure time $a$, respectively, when necessary. Let $\theta_{e}$ be the desired estimand, which must be a linear combination of the unique treatment effects in $\bm{\theta}$. Then $\theta_{e} = \bm{v}^T \bm{\theta}$ for some vector $\bm{v}$ of weights on the unique treatment effects. For example, a vector $\bm{v}$ with all zero entries except for a 1 in one entry would pick out a single unique treatment effect, and a vector $\bm{v}$ with equal entries summing to 1 would average over all unique treatment effects. In addition to averages of certain effects, $\theta_e$ could also be a difference between two treatment effects, or a more complicated linear combination, as desired. For example, heterogeneity of effects could be assessed by estimating the difference between two effects as discussed by \cite{shahn_group_2024} and \cite{xu_factorial_2024}.

Furthermore, define $\bm{F}$ as the matrix such that $\bm{F} \bm{\theta} = E[\bm{d}]$ under the specified assumption setting. Since each $E[D_{i,i',j,j'}]$ under Assumptions~\ref{as:no-spill}--\ref{as:par-trends} and a specified assumption setting is a linear combination of up to four unique treatment effects, such a matrix exists, with all entries either 0, 1, or -1.

\begin{theorem} \label{thm:FTw}
For any assumption setting with corresponding vector of unique treatment effects $\bm{\theta}$ and any target estimand $\theta_e = \bm{v}^T \bm{\theta}$, assuming that $\bm{w}$ is independent of $\bm{y}$, $\hat{\theta}$ is an unbiased estimator of $\theta_e$ if $\bm{F}^T \bm{w} = \bm{v}$.
\end{theorem}

\noindent \emph{Proof.} Let $\hat{\theta} = \bm{w}^T \bm{d}$ with $\bm{w}$ satisfying $\bm{F}^T \bm{w} = \bm{v}$. Then: \begin{align}
E[\hat{\theta}] &= E \left[ \bm{w}^T \bm{d} \right] = \bm{w}^T E[\bm{d}] = \bm{w}^T F \bm{\theta} = \bm{v}^T \bm{\theta} = \theta_e. \end{align}

The existence of a solution and, if one exists, the dimension of the space of solutions can be assessed through rank conditions.

\begin{theorem} \label{thm:rank}
Let $\bm{F}$, $\bm{w}$, $\bm{v}$, and $\bm{d}$ be as defined previously. Then the following are true about the set of estimators of the form $\hat{\theta} = \bm{w}^T \bm{d}$ that are unbiased for $\theta_e$ under the assumption setting:
\begin{itemize}
\item If $rank(\bm{F}^T|\bm{v}) > rank(\bm{F}^T)$, then there are no estimators $\hat{\theta}$ of this form that are unbiased for $\theta_e$.
\item If $rank(\bm{F}^T|\bm{v}) = rank(\bm{F}^T) = \binom{N}{2} \binom{J}{2}$, then there is a unique estimator $\hat{\theta}$ of this form that is unbiased for $\theta_e$, defined by the unique $\bm{w}$ that solves $\bm{F}^T \bm{w} = \bm{v}$.
\item If $rank(\bm{F}^T|\bm{v}) = rank(\bm{F}^T) < \binom{N}{2} \binom{J}{2}$, then there are infinitely many estimators $\hat{\theta}$ of this form that are unbiased for $\theta_e$. The dimension of unique such estimators is $(N-1)(J-1) - rank(\bm{F})$.
\end{itemize}
\end{theorem}

\noindent \emph{Proof.} The proof relies on linear algebra results on non-homogeneous systems of linear equations and the rank-nullity theorem (\cite{george_course_2024}, pp. 41 and 89, respectively). The full proof of the theorem, with two necessary lemmata, is given in Appendix~\ref{appx:Proofs}.

\subsection{Minimum variance estimator}

For settings where there are many unbiased estimators of this form, I propose using the one with the lowest estimated variance. Again, treating the weights as fixed (i.e., independent of the outcomes), the variance of the estimator for any $\bm{w}$ is given by:
\begin{equation} \label{eqn:var-plug}
Var(\hat{\theta}|\bm{w}) = Var(\bm{w}^T \bm{d}) = Var(\bm{w}^T \bm{A} \bm{y}) = \bm{w}^T \bm{A} Var(\bm{y}) \bm{A}^T \bm{w},
\end{equation}
written in terms of $Var(\bm{y})$ rather than $Var(\bm{d})$ because the correlation among different two-by-two DID estimators is not trivial. In general, $Var(\bm{y})$ will not be known \emph{a priori}. For the design and selection of the weights, then, a working covariance matrix can be used, denoted by $\bm{M}$. The working variance of the estimator is then estimated for a given $\bm{M}$ by:
\begin{equation} \label{eqn:var-full}
Var(\hat{\theta}|\bm{w},\bm{M}) = \bm{w}^T \bm{A} \bm{M} \bm{A}^T \bm{w}.
\end{equation}
Let $\bm{w}^*$ be the weight vector that minimizes $Var(\hat{\theta}|\bm{w},\bm{M})$ under the constraint $\bm{F}^T \bm{w}^* = \bm{v}$.

The variance can then be estimated in the design phase \emph{a priori} using $\bm{M}$. Since $\bm{w}^*$ is selected among weights that give unbiased estimation regardless of $\bm{M}$, misspecification of $\bm{M}$ results only in reduced efficiency, not bias. Moreover, even for variance minimization, the working covariance structure $\bm{M}$ does not need to be specified exactly, only up to a constant factor. The features of $\bm{M}$ that are necessary for the selection of the weights are the correlation within (and, if applicable, between) units across periods and the relative variances of the units and, if applicable, time periods. The working covariance matrix can be written as follows:
\begin{equation}
\bm{M} = \bm{m}_v^T \bm{M}_r \bm{m}_v,
\end{equation}
where $\bm{M}_r$ is the working correlation matrix and $\bm{m}_v$ is the vector of (relative) variances of the individual observations. If the units are independent, then $\bm{M}_r$---and thus $\bm{M}$---will be block-diagonal, with non-zero entries only for the within-unit covariances. Several common structures could then be specified for the within-unit covariances:
\begin{itemize}
\item Independence: assuming independence of observations would require $\bm{M}_r$ to be the identity matrix. This would generate a diagonal $\bm{M}$, where the diagonal entries indicate the relative variances of the different unit-period observations.
\item Exchangeable/compound symmetric: assuming compound symmetry in the within-unit correlation would require $\bm{M}_r$ to be block-diagonal with the non-zero off-diagonal entries all equal to the intra-unit (intra-cluster) correlation coefficient, $\rho$.
\item Auto-regressive, AR(1): assuming auto-regressive correlation of order 1 would require $\bm{M}_r$ to be block-diagonal with non-zero off-diagonal entries equal to the first-order correlation, $\rho$, raised to a power equal to the difference between the two periods.
\end{itemize}
If the units are exchangeable (e.g., clusters prior to randomization in a SWT), then all of the blocks will be equivalent to one another. More complex variance structures can be implied by specific outcome models as well (see, e.g., \cite{hooper_sample_2016} and \cite{kasza_impact_2019}).

\subsection{Estimation and inference}
Once the weight vector, $\bm{w}^*$, is determined, the estimator is given by:
\begin{equation}
\hat{\theta}^* = \bm{w}^{*T} \bm{d} = \bm{w}^{*T} \bm{A} \bm{y}.
\end{equation}
Inference can proceed using an appropriate plug-in estimator $\widehat{Var}(\bm{y})$ in Equation~\ref{eqn:var-plug}:
\begin{equation} \label{eqn:var-plugin}
Var(\hat{\theta}^*) = \bm{w}^{*T} \bm{A} \widehat{Var}(\bm{y}) \bm{A}^T \bm{w}^*,
\end{equation}
with a sampling distribution depending on the assumed or estimated distribution of $\bm{y}$. This would likely arise from an asymptotic sampling- or model-based approach to inference. This approach has been predominant in the observational staggered adoption literature, with recent work identifying more reasonable assumptions and approaches suitable for dependent units and smaller numbers of units; see \cite{roth_whats_2023} for a summary. Model-based approaches to inference have also been common in the randomized stepped-wedge literature, usually as some form of the linear mixed effects model (see, e.g., \cite{hussey_design_2007} and \cite{hooper_sample_2016}), with robust extensions and misspecification discussed by, e.g., \cite{hughes_robust_2020}, \cite{voldal_model_2022}, and \cite{wang_how_2024}.

Alternatively, inference can proceed non-parametrically using permutation inference. This approach has its basis both in randomization tests used in the SWT case as in \cite{wang_use_2017}, \cite{thompson_robust_2018}, and \cite{kennedyshaffer_novel_2020}, and in placebo tests in the staggered adoption case as in \cite{hagemann_placebo_2019}, \cite{mackinnon_randomization_2020}, \cite{shaikh_randomization_2021}, and \cite{roth_efficient_2023}). This design-based approach to inference is used to test a sharp null hypothesis and conditions on the observed data, using the randomization or ``as-if'' randomization of the order as the basis for hypothesis testing. Applying the same estimator to the data set under a large number of permutations of the order of adoption provides a null distribution for the estimator to which the observed estimate can be compared. In the remainder of this paper, I focus on this design-based permutation inference approach.

\subsection{Algorithm for construction of estimator} \label{sec:algo}

In summary, estimation and inference proceed by the following process.
\begin{enumerate}
\item Based on the number of periods $J$ and the number of units $N$, identify the matrix $\bm{A}$ that converts the observations $\bm{y}$ to the two-by-two DID estimators $\bm{d}$.
\item Determine the assumption setting (i.e., types of treatment heterogeneity permitted) and the target estimand $\theta_e$.
\item Based on the assumption setting and the treatment adoption sequences, identify the vector of unique treatment effects $\bm{\theta}$ and the matrix $\bm{F}$ such that $E[\bm{d}] = \bm{F} \bm{\theta}$. Identify the vector $\bm{v}$ such that $\theta_e = \bm{v}^T \bm{\theta}$.
\item Find the space of weight vectors $\bm{w}$ such that $\bm{F}^T \bm{w} = \bm{v}$ and, if relevant, restrict to the set of weights that give unique estimators.
\item Determine an appropriate working covariance matrix $\bm{M}$ (up to a scalar) using subject-matter knowledge, pilot study data, or an assumed outcome model.
\item Find the weight vector $\bm{w}^*$ among the unbiased solutions $\bm{w}$ that minimizes the working variance $\bm{w}^T \bm{A} \bm{M} \bm{A}^T \bm{w}$.
\item Once observed data $\bm{y}$ are obtained, the estimator is then given by $\hat{\theta}^* = \bm{w}^{*T} \bm{A} \bm{y}$.
\item The estimated variance can then be found by $\widehat{Var}(\hat{\theta}^*) = \bm{w}^{*T} \bm{A} \widehat{Var}(\bm{y}) \bm{A}^T \bm{w}^*$, using a plug-in estimator $\widehat{Var}(\bm{y})$. Alternatively, permutation-based inference can be conducted.
\end{enumerate}

See \url{http://github.com/leekshaffer/GenDID} for \texttt{R} code implementing this algorithm.

\section{Toy Example} \label{sec:toy}
To illustrate the algorithm and the approach described herein, consider first a setting with $N=2$ units and $J=3$ periods. The first unit adopts the treatment starting in the second period and the second unit adopts the treatment starting in the third period (i.e., $T_1 = 2$ and $T_2 = 3$). There are thus three two-by-two DID estimators:
\begin{align*}
D_{1,2,1,2} &= (Y_{12} - Y_{11}) - (Y_{22} - Y_{21}) \mbox{,  (type 2)} \\
D_{1,2,1,3} &= (Y_{13} - Y_{11}) - (Y_{23} - Y_{21}) \mbox{,  (type 4)} \\
D_{1,2,2,3} &= (Y_{13} - Y_{12}) - (Y_{23} - Y_{22}) \mbox{,  (type 5)} \\
\end{align*}
The matrix relating the vector of estimators $\bm{d} = \begin{pmatrix} D_{1,2,1,2} & D_{1,2,1,3} & D_{1,2,2,3} \end{pmatrix}^T$ to the vector of outcomes $\bm{y} = \begin{pmatrix} Y_{11} & Y_{12} & Y_{13} & Y_{21} & Y_{22} & Y_{23} \end{pmatrix}^T$ is given by:
\begin{equation}
\bm{A} = \begin{pmatrix} -1 & 1 & 0 & 1 & -1 & 0 \\ -1 & 0 & 1 & 1 & 0 & -1 \\ 0 & -1 & 1 & 0 & 1 & -1 \end{pmatrix},
\end{equation}
which has rank 2.

I consider several different assumption settings. Throughout, the DID building blocks ensure that weights on rows and columns sum to 0, as discussed by \cite{matthews_stepped_2017}. 

\subsection{Assumption S5: Homogeneity}
Assuming full treatment effect homogeneity, there is a single treatment effect ($\bm{\theta} = \theta$) and the expectation of the estimators is given by:
\begin{equation}
E[\bm{d}] = \begin{pmatrix} \theta \\ 0 \\ -\theta \end{pmatrix} = \begin{pmatrix} 1 \\ 0 \\ -1 \end{pmatrix} \theta \equiv \bm{F} \bm{\theta}.
\end{equation}
There is only one target estimand of natural interest here so $\bm{v} = 1$, yielding $\theta_e = \theta$. The ranks of $\bm{F}^T$ and $\left( \bm{F}^T | \bm{v} \right) = \begin{pmatrix} 1 & 0 & -1 & 1 \end{pmatrix}$ are both 1. By Theorem~\ref{thm:rank}, then, the space of unique estimators of this form is one-dimensional.
The solution space to $\bm{F}^T \bm{w} = \bm{v}$ is the space of vectors of the form $\begin{pmatrix} x & y & x-1 \end{pmatrix}^T$ for any real values $x,y$. One dimension is reduced, however, when right-multiplied by $\bm{A}$, which yields:
\[ \bm{w}^T \bm{A} = \begin{pmatrix} -x-y & 1 & x+y-1 & x+y & -1 & -x-y+1 \end{pmatrix}, \]
which only depends on the single free parameter $x+y$. A few special cases can be seen for specific values of $x+y$:
\begin{itemize}
\item If $x+y=1$, then the estimator is simply the type-2 estimator $D_{1,2,1,2}$, the ``clean'' comparison as described by \cite{goodman-bacon_difference--differences_2021} or the ``crossover'' estimator CO-1 proposed by \cite{kennedyshaffer_novel_2020};
\item If $x+y=0$, then the estimator uses only the comparison of unit two adopting the treatment while unit one is always treated; and
\item If $x+y=1/2$, then the estimator is ``centrosymmetrized'' as described by \cite{bowden_centrosymmetry_2021}, using both of the switches equally, equivalent to the horizontal row-column estimator proposed by \cite{matthews_stepped_2017} or the ``crossover'' estimator CO-3 that uses always-treated units as controls in \cite{kennedyshaffer_novel_2020}.
\end{itemize}
Note that the purely vertical estimators described by \cite{matthews_stepped_2017} and \cite{thompson_robust_2018} cannot be constructed as a weighted average of the two-by-two DID estimators.

Assuming independent and homoskedastic observations, $\bm{M} = \bm{I}_6$, the six-by-six identity matrix. The constrained minimization of the variance, then, gives the solution $x+y=1/2$ and the estimator $\hat{\theta}^* = \frac{D_{1,2,1,2} - D_{1,2,2,3}}{2}$. As mentioned, this corresponds to the ``centrosymmetrized'' estimator discussed by \cite{matthews_stepped_2017} and \cite{bowden_centrosymmetry_2021}; the latter proves that this estimator has minimum variance under these assumptions. The same estimator minimizes the variance under a working correlation assumption that is compound symmetric or AR(1), regardless of the correlation value, as long as units are exchangeable and independent.

\subsection{Assumption S4: Calendar-Time Heterogeneity}
If there is treatment effect heterogeneity only by calendar-time, then there are two unique treatment effects and $\bm{\theta} = \begin{pmatrix} \theta^{(4)}_2 \\ \theta^{(4)}_3 \end{pmatrix}$, where $\theta^{(4)}_j$ is the effect of treatment in period $j$. Now, $\bm{F} = \begin{pmatrix} 1 & 0 \\ 0 & 0 \\ -1 & 0 \end{pmatrix}$. Since the last column only has zero entries, it is impossible to estimate $\theta^{(4)}_3$. This occurs since no DID comparisons here compare a treated and untreated unit in period 3. Formally, any $\bm{v}$ that does not have a 0 as the second entry will yield $2 = rank(\bm{F}^T|\bm{v}) > rank(\bm{F}^T) = 1$. There is a one-dimensional space of unbiased estimators of $\theta^{(4)}_2$, however, given by weight vectors of the form $\begin{pmatrix} x & 0 & x-1 \end{pmatrix}^T$ for real values $x$; these are similar to those in the previous subsection, but forbidding any non-zero weight on $D_{1,2,1,3}$.

\subsection{Assumption S3: Exposure-Time Heterogeneity}
If there is treatment effect heterogeneity only by exposure-time (time on treatment), then there are two unique treatment effects and $\bm{\theta} = \begin{pmatrix} \theta^{(3)}_1 \\ \theta^{(3)}_2 \end{pmatrix}$, where $\theta^{(3)}_a$ is the effect of treatment in the $a$th treatment period. Now, $\bm{F} = \begin{pmatrix} 1 & 0 \\ -1 & 1 \\ -2 & 1 \end{pmatrix}$. This is a full-rank matrix, so any linear combination of $\bm{\theta}$ can be estimated without bias. Since $rank(\bm{F}^T) = 2 = rank(\bm{A}^T)$, all solutions are unique (in terms of weights on the observations and thus the estimator itself).

Letting $\bm{v} = \begin{pmatrix} 1/2 \\ 1/2 \end{pmatrix}$, so that $\theta_e = \frac{1}{2} \theta^{(3)}_1 + \frac{1}{2} \theta^{(3)}_2$, the simple average of the two treatment effects, unbiased estimators have weights of the form $\bm{w}^T = \begin{pmatrix} 1+y & \frac{1}{2} - y & y \end{pmatrix}$, which gives observation weights $\bm{w}^T \bm{A} =\begin{pmatrix} - \frac{3}{2} & 1 & \frac{1}{2} & \frac{3}{2} & -1 & -\frac{1}{2} \end{pmatrix}$. In other words, all unbiased estimators of this form are equivalent to $\hat{\theta} = D_{1,2,1,2} + \frac{1}{2} D_{1,2,1,3}$.

Letting $\bm{v} = \begin{pmatrix} 1 \\ 0 \end{pmatrix}$, so that $\theta_e = \theta^{(3)}_1$, the first treated period effect, unbiased estimators have weights of the form $\bm{w}^T = \begin{pmatrix} 1+y & -y & y \end{pmatrix}$, which gives observation weights $\bm{w}^T \bm{A} = \begin{pmatrix} -1 & 1 & 0 & 1 & -1 & 0 \end{pmatrix}$. In other words, all unbiased estimates of this form are equivalent to $\hat{\theta} = D_{1,2,1,2}$. Again, this is equivalent to the ``crossover'' estimator CO-1 from \cite{kennedyshaffer_novel_2020}, which cannot be centrosymmetrized according to \cite{bowden_centrosymmetry_2021} because there is treatment effect heterogeneity. Similarly, \cite{goodman-bacon_difference--differences_2021} clarifies that the comparison of the switching unit 2 to the always-treated unit 1 between periods 2 and 3 is not unbiased for the first-period effect.

Note that the results in this assumption setting do not depend on a working covariance matrix, since the space of unbiased solutions is of dimension zero. In general, this will only occur with few clusters and/or few assumed homogeneities.

\section{Comparisons to Existing Estimators} \label{sec:Comps}

Several existing estimators can be considered under this framework as well, as either weighted averages of certain two-by-two comparisons or, more generally, weighted averages of observations. \cite{goodman-bacon_difference--differences_2021} discusses this in general for the two-way fixed effects model in the non-randomized staggered adoption setting, while \cite{matthews_stepped_2017} and \cite{lindner_heterogeneous_2021} discuss it for the stepped-wedge setting. Here, I consider several alternative estimators that have been proposed for this setting and their relationship to the generalized estimator. The first group were designed for the staggered adoption setting- (see \cite{arkhangelsky_causal_2024} for a summary), the second group specifically target the effect in the first period on treatment, and the last estimator uses only vertical comparisons. Comparisons can also be made to methods designed to assess heterogeneity using DID approaches as in \cite{shahn_group_2024} and \cite{xu_factorial_2024}.

\subsection{Staggered Adoption Methods}

Until recently, staggered adoption panel data analyses often used the two-way fixed-effects model (TW). \cite{goodman-bacon_difference--differences_2021} demonstrated that this could be expressed as a weighted average of two-by-two DID comparisons. Considering each pair of units and periods as a unique comparison (as opposed to the Goodman-Bacon decomposition itself, which combines across timing groups), this method equally weights all comparisons where one unit switches and the other does not, with negative weights for those where the earlier-switching unit is always-treated and the later-switching unit switches. This implicitly makes assumption S5, full treatment effect homogeneity. The TW method is equivalent to the generalized DID with assumption S5 under the assumption that observations are independent and homoskedastic, giving a highly efficient, symmetrized estimator.

\cite{callaway_difference--differences_2021} propose a method using only ``clean'' two-by-two comparisons (CS); i.e., restricting to comparisons where unit $i'$ is untreated in both periods $j$ and $j'$ (Types 1--3 in Table~\ref{tbl:2x2-cats}). This can be further restricted to always comparing to a set of never-treated or last-treated control units, but we consider the more general version here. Moreover, they only include comparisons where the baseline period is the last untreated period of unit $i$; i.e., two-by-two components $D_{i,i',T_i-1,j'}$ where $T_i \le j' < T_{i'}$. These are combined by equal weights into group-time ATTs (where group is defined as the set of units with the same treatment adoption time), which can then be weighted to the desired estimand. Four proposed summary estimates are the ``simple'' ATT (averaging across group-times proportional to the number of units in each group), the ``dynamic'' ATT (averaging by time-on-treatment and then weighting equally across those exposure times), the ``group'' ATT (averaging within groups, and then equally across groups), and the ``calendar'' ATT (averaging within calendar time periods, and then equally across periods). These all imply different weights on the included two-by-two DID estimators.

\cite{sun_estimating_2021} similarly propose a method using only ``clean'' comparisons (SA), specifically comparisons against the never-treated or last-treated group where the last untreated period of unit $i$ is used as the baseline; i.e., two-by-two components $D_{i,i',T_i-1,j'}$, where $j' > T_i - 1$ and $i' \in C$, the set of units with the latest treatment adoption time ($T_{i'} = \max_i T_i$ for $i' \in C$). These are combined, weighting equally across units within each time period $j'$, and then equally weighted across periods. Further restricting the comparison set limits the two-by-two DID estimators that get non-zero weights.

The above methods target a specific estimand by both limiting the comparisons to only allow comparisons with an expectation equal to the desired group-time treatment effect (allowing heterogeneity by both calendar time and exposure time) and then by weighting these together in particular ways. Both can be expressed as weighted averages of two-by-two DID components, as in the generalized DID framework proposed here. Identifying the same target effects can be done more efficiently, however, by incorporating additional components that do not change the expectation but can reduce variance. The tradeoff for this more general approach is that a stronger parallel trends assumption may be required, as more cluster-periods are used in the estimation.

\subsection{First-Period Effect Methods}

The method proposed by \cite{de_chaisemartin_two-way_2020} (CH) allows comparisons to not-yet-treated groups, but only one-period ahead differences; i.e. only Type 2 comparisons $D_{i,i',T_i-1,T_i}$, where $T_{i'} > T_i$. These are then averaged within each timing group, and then across groups, weighted equally. This inherently restricts the target estimand to the effect of treatment in the first period of adoption.

Similarly, the crossover estimator proposed by \cite{kennedyshaffer_novel_2020} only considers first-period effects. CO-1 weights equally across timing groups, and is thus equal to the \cite{de_chaisemartin_two-way_2020} estimator. CO-2 uses the same comparisons, but different weights, weighting each timing group proportional to the harmonic mean of the number of treated and control units. CO-3 is similar, but allows always-treated units to act as controls as well, incorporating Type 5 comparisons as well. Weighting is similar to CO-1. Again, these all target the first-period effect, but with different weights having different properties under calendar-time or unit heterogeneity (CO-3 in particular requires no dynamic effects).

These two methods target a very specific estimand, the first-period effect of adoption, and are best suited to ignoring or averaging over calendar time heterogeneity. CO-3 has improved efficiency under the assumption of no dynamic treatment effects because it uses more data (see, e.g., \cite{bowden_centrosymmetry_2021}), but can become biased for the first-period effect if that assumption does not hold (\cite{kennedyshaffer_novel_2020}). The generalized DID method also can be designed to target this effect alone; it will generally use more comparisons and be more efficient under any specific assumption setting. However, it loses the simplicity of restricting the set to just include those comparisons that clearly target that specific effect which in non-randomized settings may require a stronger parallel trends assumption.

\subsection{Observation Weight Method}

Finally, the non-parametric within-period method (NP) proposed by \cite{thompson_robust_2018} uses individual observations as the components instead of two-by-two DID comparisons. It compares treated to untreated units within each calendar period, and then averages these comparisons across periods, with weights either equal across periods, proportional to the number of treated groups, or inversely proportional to the estimated variance of the comparison. For the inverse-variance weights, we assume equal variance across observations rather than using the sample variance, so that weights can be computed without the observation data. The within-period approach is particularly useful when calendar-time heterogeneity is expected to be more pronounced or of greater interest than exposure-time heterogeneity. However, the lack of horizontal comparisons limits the power and efficiency of the method (see, e.g., \cite{matthews_stepped_2017}, \cite{thompson_robust_2018}, and \cite{kennedyshaffer_novel_2020}).

\subsection{Comparisons}

Observation weights and, for all except the final method described, DID weights for these methods are implemented in the \verb|Comp_Ests_Weights| function in the R code, to enable investigators to compare the weights used in these different approaches to those under the generalized DID framework. These weights can all be calculated \emph{a priori}, so that the advantages and disadvantages of the weighting schemes can be considered without being influenced by the observed data. Overall, the generalized DID method tends to use more information to target similar effects and allows more flexibility for the researcher. This may result in reduced variance and improved efficiency of estimation, especially under more effect homogeneity and if the working covariance matrix is correctly specified. The tradeoff, however, is that it is generally a more complex estimator than the ones proposed here, with less clearly interpretable weights. By using more comparisons, it also requires the assumptions made (e.g., parallel trends and any homogeneity assumptions) to hold true for a broader set of units and periods.

\section{Data Example: Stepped Wedge Trial} \label{sec:SWTex}

To illustrate the uses of this approach, consider the stepped wedge trial conducted in 2012 and reported by \cite{durovni_impact_2014} and \cite{trajman_impact_2015}. I use the outcomes reported by \cite{trajman_impact_2015}, assessing the impact on treatment outcomes for patients diagnosed with tuberculosis. Under the control condition, patients were diagnosed by sputum smear examination, while under the intervention condition patients were diagnosed with the XpertMTB/RIF test, hypothesized to be faster and more sensitive in confirming diagnosis and, in particular, diagnosing rifampicin-resistant tuberculosis.

This study randomized fourteen clusters into seven treatment sequences, with observations in each of eight months. The first sequence, consisting of two clusters, began the intervention in the second month; each subsequent month, two more clusters began the intervention (\cite{durovni_impact_2014}). The patient outcomes, namely unsuccessful treatment of the patient defined as any outcome other than ``cure and treatment completion without evidence of failure'' (\cite{trajman_impact_2015}), were retrieved using the Brazilian national tuberculosis information system over one year after diagnosis.

The original analysis found a reduction in this composite outcome, although it was not statistically significant: an odds ratio of 0.93 with a 95\% confidence interval of 0.79--1.08 (\cite{trajman_impact_2015}). Accounting for time trends in the analyses, however, re-analysis by \cite{thompson_robust_2018} using the purely vertical non-parametric within-period method found a larger and statistically significant effect: an odds ratio of 0.78 with a p-value of 0.02. Similar results were found by \cite{kennedyshaffer_novel_2020} using the crossover method, which estimated an odds ratio of 0.70 with a permutation-based p-value of 0.046. These results indicate the possible impacts of different treatment effect heterogeneity assumptions and different choices of estimand.

I re-analyze these data using several estimators from the generalized DID framework, beginning with four overall treatment effects:
\begin{itemize}
\item $\hat{\theta}^{(5)}$, the estimate under treatment effect homogeneity (assumption S5);
\item $\frac{1}{6} \sum_{j=2}^7 \hat{\theta}^{(4)}_j$, the average period-specific treatment effect across months 2--7 of the trial under calendar-time heterogeneity (assumption S4);
\item $\frac{1}{7} \sum_{a=1}^7 \hat{\theta}^{(3)}_a$, the average time-on-treatment effect across exposure times 1--7 under exposure-time heterogeneity (assumption S3); and 
\item $\frac{1}{21} \sum_{j=2}^7 \sum_{a=1}^{j-1} \hat{\theta}^{(2)}_{j,a}$, the average across all identifiable calendar- and exposure-time combinations of the trial under both heterogeneities (assumption S2).
\end{itemize}
Note that the period-specific treatment effects in month 8 (the last month of the study) are not identifiable since all clusters were in the treated condition at that point. These clusters do contribute to estimation of $\hat{\theta}^{(5)}$ and $\hat{\theta}^{(3)}_7$, however, under assumptions S5 and S3, as those assumptions allow us to borrow information across different calendar times in estimating common treatment effects. The odds ratio (calculated using log-odds of events in each cluster-period as the outcome) and risk difference (calculated using proportion of events in each cluster-period as the outcome) estimates and permutation-test p-values (using 1000 permutations) are shown in Table~\ref{tbl:res1}. All results shown were calculated using an exchangeable (within-cluster) variance structure with intracluster correlation coefficient $\rho = 0.003$ as estimated by \cite{thompson_robust_2018}. Results under independent and auto-regressive variance structures are quite similar and shown in Appendix~\ref{appx:extrares}.

\begin{table}[!ht]
\caption{Odds ratio and risk difference estimates and permutation test p-values for generalized DID estimators of summary effects of XpertMTB/RIF testing on tuberculosis outcomes in Brazil, 2012. \label{tbl:res1}}
\begin{center}
\begin{tabular}{ll|r|r|r|r|}
& & \multicolumn{2}{c|}{Odds Ratio} & \multicolumn{2}{c|}{Risk Difference} \\
Assumption & Estimator & Estimate & P-Value  & Estimate & P-Value \\ \hline
S5 & $\hat{\theta}^{(5)}$ & 0.771 & 0.013 & -7.35\% & 0.020\\
S4 & $\frac{1}{6} \sum_{j=2}^7 \hat{\theta}^{(4)}_j$ & 0.784 & 0.021 &-6.81\% & 0.054 \\
S3 & $\frac{1}{7} \sum_{a=1}^7 \hat{\theta}^{(3)}_a$ & 0.828 & 0.413 & -5.57\% & 0.456 \\
S2 & $\frac{1}{21} \sum_{j=2}^7 \sum_{a=1}^{j-1} \hat{\theta}^{(2)}_{j,a}$ & 0.801 &  0.149 & -6.11\% &  0.229 \\
\end{tabular}
\end{center}
\end{table}

Using the (scaled) variance calculated using Equation \ref{eqn:var-full} also allows a comparison of the relative efficiency of these estimators, even without a specific variance estimate. Under this exchangeable variance structure, the relative efficiencies are 1.05, 2.76, and 1.77, comparing the S4, S3, and S2 estimators, respectively, to the S5 estimator $\hat{\theta}^{(5)}$. This quantifies the variance trade-off that occurs in order to gain the robustness to bias and target a specified estimand under different forms of heterogeneity.

\begin{figure}[!ht]
\begin{center}
\includegraphics[width=5in]{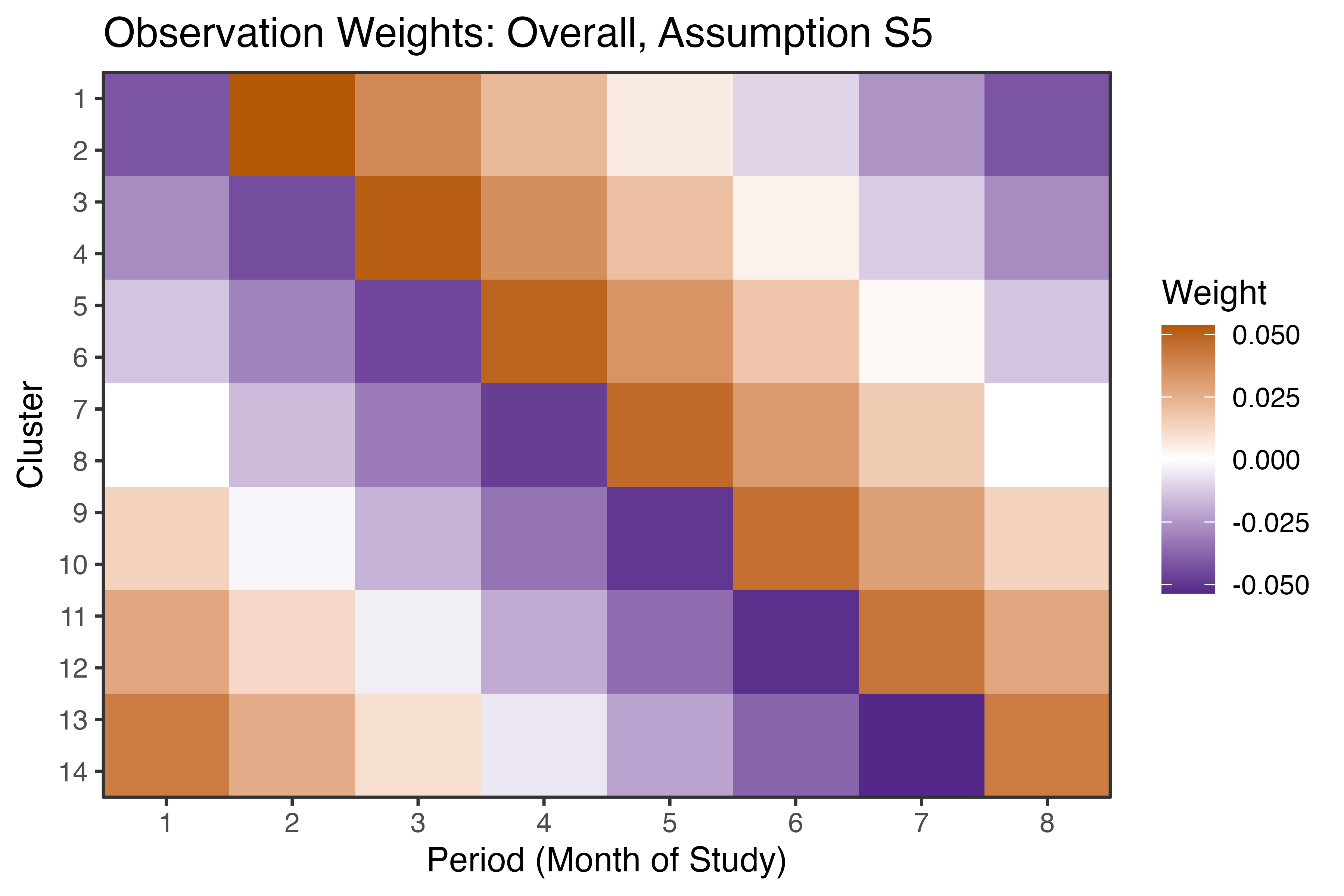}
\end{center}
\caption{Weights on cluster-period observations for the estimator $\hat{\theta}^{(5)}$ fitted under an exchangeable variance structure with $\rho = 0.003$. \label{fig:obsW}}
\end{figure}

To understand the form of the estimator, we can also examine the weights it gives to each observation (note: the weights on each two-by-two DID estimator are also available, but less interpretable as many different weight vectors can yield the same weights on the observations). The weights for the estimator $\hat{\theta}^{(5)}$ are shown in the heat map in Figure~\ref{fig:obsW}. Pairs of clusters (e.g., 1 and 2) have nearly identical weights because they are in the same sequence and assumed to be exchangeable. These weights can be compared to the information content of specific cells under different heterogeneity and correlation assumptions as proposed by \cite{kasza_information_2019} and \cite{kasza_information_2019-1}. Note heuristically that in this case, the largest (in magnitude) weights are near the diagonal, followed by the top-right and bottom-left cells, which are also the highest information-content cells as shown in Figure 4 of \cite{kasza_information_2019}. In addition, the weights satisfy the symmetry discussed by \cite{matthews_stepped_2017} and \cite{bowden_centrosymmetry_2021}, which maximizes efficiency. Slight differences arise when assumption S4 is used instead with the calendar time-averaged estimator; Figure~\ref{fig:obsWS4} displays these weights.

This method provides flexibility to investigate other treatment effects in the heterogeneous settings as well. Considering calendar-time effects under assumption S4, for example, one can estimate the treatment effect in each of months 2 through 7, $\hat{\theta}^{(4)}_2, \ldots, \hat{\theta}^{(4)}_7$. For additional robustness against exposure-time heterogeneity, one can also estimate month-specific treatment effects under assumption S2 by averaging across exposure times within each of those months: $\hat{\theta}^{(2)}_j \equiv \frac{1}{j-1} \sum_{a=1}^{j-1} \hat{\theta}^{(2)}_{a,j}$ for all $j=2,\ldots,7$. The estimation and permutation test results on both the multiplicative (odds ratio) and additive (risk difference) scales are shown in Table~\ref{tbl:res2}. These results can be compared to the results in Table 2 of \cite{thompson_robust_2018}, which reports period-specific risk differences estimated using the purely vertical non-parametric within-period estimator (note that those are labelled by treatment periods, one less than the corresponding month numbers given here). Differences between the methods arise as the results presented here use horizontal information as well, but target the same estimand. Note that no p-values are adjusted for multiple testing.

\begin{table}[!ht]
\caption{Odds ratio and risk difference estimates and permutation test p-values for generalized DID estimators of calendar-time treatment effects of XpertMTB/RIF testing on tuberculosis outcomes in Brazil, 2012, using an exchangeable variance structure with ICC 0.003. \label{tbl:res2}}
\begin{center}
\begin{tabular}{l|r|r|r|r|r|r|r|r|}
& \multicolumn{4}{c|}{Odds Ratio} & \multicolumn{4}{c|}{Risk Difference} \\
 & \multicolumn{2}{c|}{Assumption S4} & \multicolumn{2}{c|}{Assumption S2} & \multicolumn{2}{c|}{Assumption S4} & \multicolumn{2}{c|}{Assumption S2} \\
Month & Estimate & P-Value & Estimate & P-Value & Estimate & P-Value & Estimate & P-Value \\ \hline
2 & 0.903 & 0.692 & 0.808 & 0.414 & -2.38\% & 0.790 & -4.01\% & 0.543\\
3 & 0.758 & 0.369 & 0.810 & 0.527 & -3.97\% & 0.514 & -1.73\% & 0.808\\
4 & 0.683 & 0.008 & 0.736 & 0.047 & -11.64\% & 0.012 & -8.65\% & 0.084\\
5 & 0.832 & 0.193 & 0.902 & 0.595 & -6.19\% & 0.150 & -3.24\% & 0.578\\
6 & 0.716 & 0.062 & 0.750 & 0.171 & -10.93\% & 0.030 & -9.02\% & 0.157\\
7 & 0.835 & 0.518 & 0.812 & 0.516 & -5.75\% & 0.635 & -6.15\% & 0.636\\
\end{tabular}
\end{center}
\end{table}

We can also compare these results with those from the existing methods described in Section~\ref{sec:Comps}. These are summarized in Table~\ref{tbl:xpert-comps}.

\begin{table}[!ht]
\caption{Risk difference estimates and permutation test p-values for various staggered adoption methods and comparable generalized DID estimates for XpertMTB/RIF testing on tuberculosis outcomes in Brazil, 2012. Note: generalized DID estimators are for the most comparable estimand and assumptions, using an exchangeable correlation structure with ICC 0.003.\label{tbl:xpert-comps}}
\begin{center}
\begin{tabular}{|l|rr|l|rr|}
\multicolumn{3}{c|}{Comparison Method} & \multicolumn{3}{c|}{Generalized DID} \\
Method & Estimate & P-Value & Estimator & Estimate & P-Value \\ \hline
TW & -7.35\% & 0.018 & $\hat{\theta}^{(5)}$ & -7.35\% & 0.020 \\
CS (simple ATT) & -6.41\% & 0.214 & $\frac{1}{21} \sum_{j=2}^7 \sum_{a=1}^{j-1} \hat{\theta}_{j,a}^{(2)}$ & -6.11\% & 0.229 \\
CS (dynamic ATT) & -5.39\% & 0.472 & $\frac{1}{7} \sum_{a=1}^7 \hat{\theta}_a^{(3)}$ & -5.57\% & 0.456 \\
CS (group ATT) & -7.10\% & 0.195 & $\frac{1}{6} \sum_{a=1}^6 \sum_{j=a+1}^7 \frac{1}{7-j+a} \hat{\theta}_{j,a}^{(2)}$ & -8.38\% & 0.130 \\
CS (calendar ATT) & -5.63\% & 0.194 & $\frac{1}{6} \sum_{j=2}^7 \hat{\theta}_j^{(4)}$ & -6.81\% & 0.054 \\
SA (ATT) & -10.64\% & 0.072 & $\frac{1}{21} \sum_{j=2}^7 \sum_{a=1}^{j-1} \hat{\theta}_{j,a}^{(2)}$ & -6.11\% & 0.229 \\
CH & -7.34\% & 0.055 & $\frac{1}{6} \sum_{j=2}^7 \hat{\theta}_{j,1}^{(2)}$ & -8.41\% & 0.015 \\
CO-1 & -7.34\% & 0.055 & $\frac{1}{6} \sum_{j=2}^7 \hat{\theta}_{j,1}^{(2)}$ & -8.41\% & 0.015 \\
CO-2 & -6.97\% & 0.051 & $\hat{\theta}_1^{(3)}$ & -6.80\% & 0.043 \\
CO-3 & -7.00\% & 0.043 & $\hat{\theta}^{(5)}$ & -7.35\% & 0.020 \\
NP (time-averaged) & -4.06\% & 0.185 & - & - & - \\
NP (ATT) & -4.86\% & 0.214 & - & - & - \\
NP (inverse-variance) & -4.64\% & 0.067 & - & - & - \\ \hline
\end{tabular}
\end{center}
\end{table}

\section{Data Example: Non-Randomized Staggered Adoption} \label{sec:SAex}

In addition to using these methods in the randomized stepped-wedge trial case, we can use them to analyze observational (non-randomized) staggered adoption settings. Doing so, however, requires assessing additional assumptions and considering how the effect heterogeneity and variance assumptions comport with the assumptions necessary to interpret statistical parameters as causal parameters (i.e., Assumptions~\ref{as:no-spill}--\ref{as:par-trends}), especially the parallel trends assumption. The trade-offs of different approaches must be considered in this larger context for observational settings.

Several U.S. states implemented lottery-type financial incentives for COVID-19 vaccination in the spring and summer of 2021 to encourage uptake of the vaccines that had recently become available. Studies examining the effects of these policies using a variety of quasi-experimental methods generally found small positive effects, often not statistically significant for a single state (see, e.g., \cite{sehgal_impact_2021}, \cite{brehm_ohio_2022}, \cite{fuller_assessing_2022}, \cite{thirumurthy_association_2022}, \cite{kim_did_2023}, and \cite{lang_did_2023}). For example, \cite{brehm_ohio_2022} identified a modest positive impact of the Ohio vaccine lottery using both DID and synthetic control methods, and \cite{fuller_assessing_2022} found that it increased uptake of the first dose of the vaccine, but not of the complete series.

I re-evaluate the impact of these policies in Midwestern states, using data collected and made publicly available by \cite{fuller_assessing_2022}. I restrict the analysis to the twelve states within the Midwest geographic region delineated by the U.S. Centers for Disease Control and Prevention (\cite{national_center_for_health_statistics_geographic_2024}). The outcome variable is the cumulative percentage of the adult (18+ years old) population who received at least one dose of any COVID-19 vaccine, recorded weekly for the weeks ending May 15 through July 31, 2021 (MMWR weeks 15--30, according to \cite{morbidity_and_mortality_weekly_report_weeks_2019}). These restrictions are made to both increase the plausibility of the parallel trends assumption (by restricting to more geographically similar states), reduce the number of concurrent interventions, and focus on shorter-term impacts of these policies. Per \cite{fuller_assessing_2022}, four of these twelve states enacted lotteries during this time period: Ohio (MMWR Week 19), Illinois (24), Michigan (26), and Missouri (29). Note that lotteries are considered to be in effect starting in the week the policy was announced. Depending on the exact date, this may lead to a first week of intervention with fewer than seven days affected by the policy. Since the focus here is on the proposed estimators, I do not fully assess the necessary assumptions, but refer interested readers to the cited studies. Results here should be interpreted with caution.

Since the details of the lottery differ by state, as do the background conditions of the pandemic and vaccine uptake, it is reasonable to expect cluster-level heterogeneity, as well as both calendar- and exposure-time heterogeneity of treatment effects. Since calendar- and exposure-time heterogeneity imply distinct treatment effects for each treated cluster-period in this setting (because each cluster has a different adoption week), we can use assumption S2 to capture these heterogeneities. In addition, since the outcome is cumulative, an auto-regressive structure is the most reasonable, so an AR(1) correlation structure is used with first-difference parameter $\rho = 0.95$, estimated from the states in other regions in those same time periods.

\begin{figure}[!ht]
\begin{center}
\includegraphics[width=3.5in]{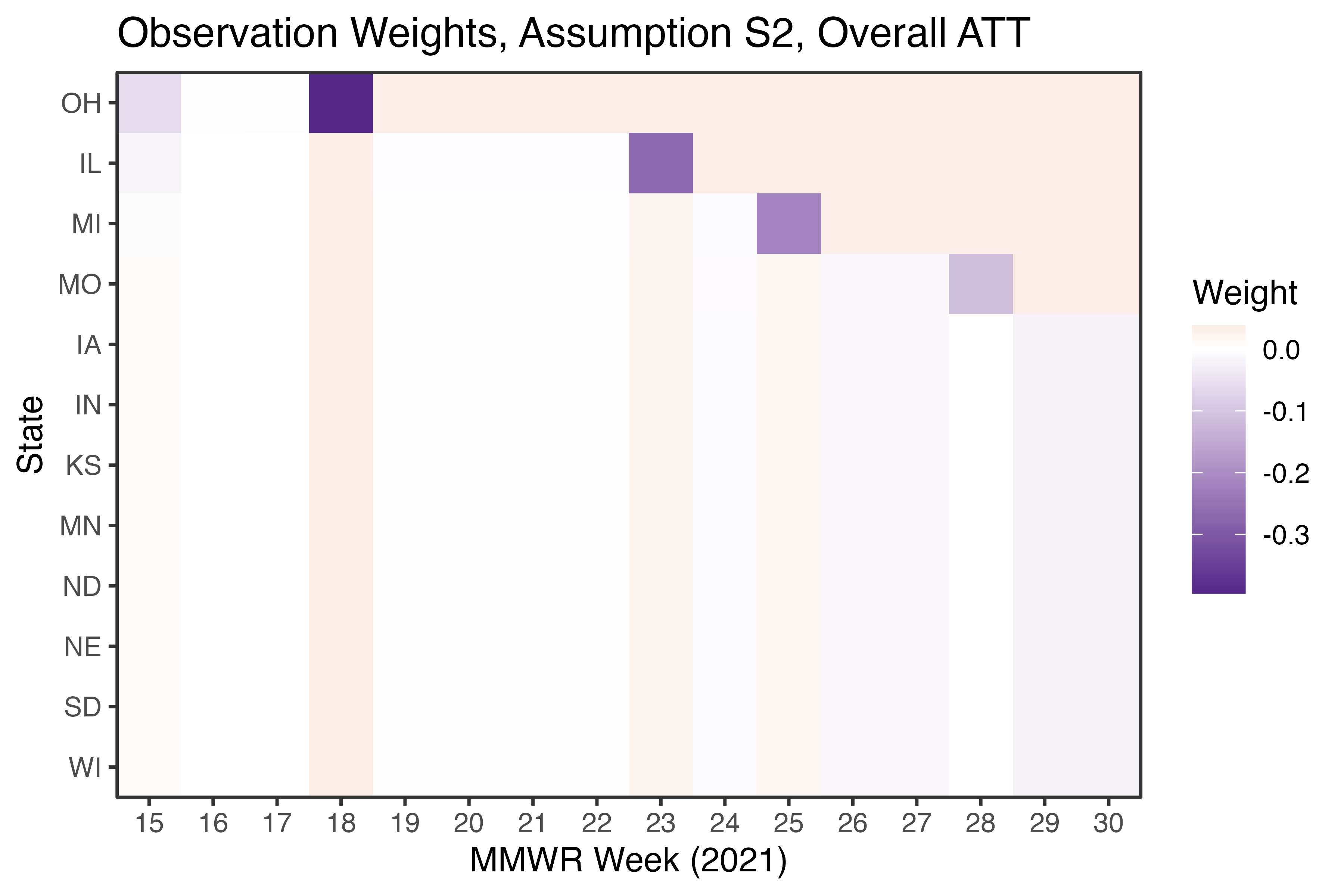}~
\includegraphics[width=3.5in]{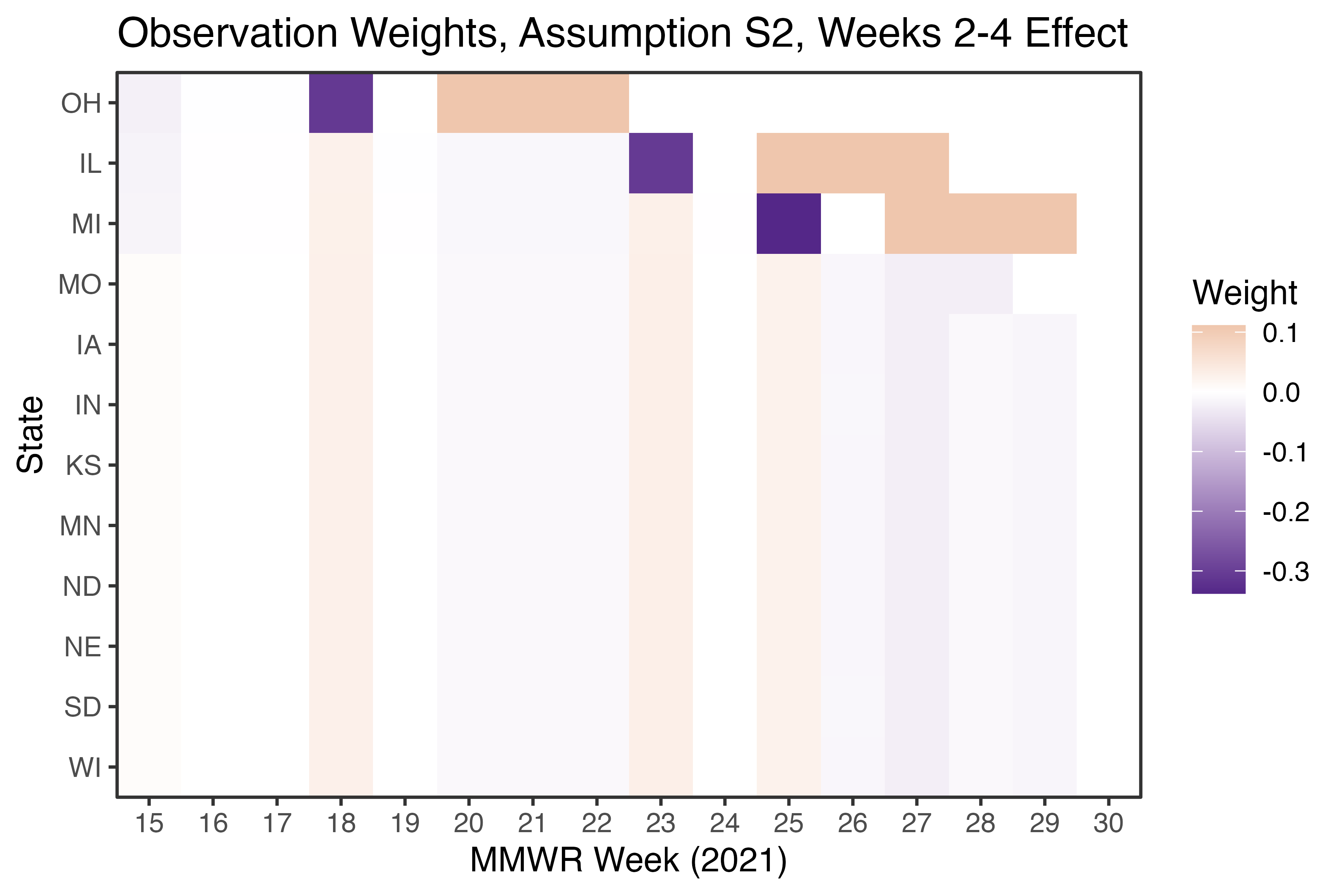}
\end{center}
\caption{Weights on cluster-period observations for the overall ATT estimator (top) and average 2--4-week effect estimator (bottom) under assumption S2 fitted under an AR(1) working correlation structure with parameter $\rho = 0.95$. \label{fig:vaxW}}
\end{figure}

Table~\ref{tbl:vax1} displays the estimates and permutation test p-values for several summary estimands of potential interest in this setting. The weightings used for the overall ATT estimator (equal weight across all of the effects) and for the average 2--4-week effect (equal weight across the treatment effects in adoption weeks 2--4 in the three states with at least four weeks of treatment) are shown in Figure~\ref{fig:vaxW}. Note that the staircase pattern of weights in the SWT example is not present here because the adoptions are not regularly spaced, but we can still identify which observations in the treated and untreated clusters are contributing the most.

\begin{table}[!ht]
\caption{Cumulative percentage difference estimates and permutation test p-values for generalized DID estimators of summary effects of vaccine lotteries on COVID-19 vaccine uptake, United States Midwest region, May--July 2021. Estimation occurred using an AR(1) working correlation structure with parameter $\rho = 0.95$.\label{tbl:vax1}}
\begin{center}
\begin{tabular}{ll|r|r|}
Description & Estimator & Estimate & P-Value  \\ \hline
Overall ATT & $\frac{1}{26} \sum_{\{j,a\}} \hat{\theta}_{j,a}^{(2)}$ & 0.537 & 0.439 \\
First-Week Effect & $\frac{1}{4} \sum_{j} \hat{\theta}_{j,1}^{(2)}$ & 0.285 & 0.155 \\
Second-Week Effect & $\frac{1}{4} \sum_{j} \hat{\theta}_{j,2}^{(2)}$ & 0.605 & 0.065 \\
Average Four-Week Effect & $\frac{1}{12} \sum_{\{j,a:~j-a \le 26, 1 \le a \le 4\}} \hat{\theta}_{j,a}^{(2)}$ & 0.483 & 0.275 \\
Average 2--4-Week Effect & $\frac{1}{9} \sum_{\{j,a:~j-a \le 26, 2 \le a \le 4\}} \hat{\theta}_{j,a}^{(2)}$ & 0.561 & 0.276 \\
State-Averaged Effect & $\frac{1}{4} \sum_{t=15}^{30} \frac{1}{31-t} \sum_{a=1}^{31-t} \hat{\theta}_{a+t,a}^{(2)}$ & 0.612 & 0.250 \\
Ohio Effect & $\frac{1}{12} \sum_{a=1}^{12} \hat{\theta}_{a+18,a}^{(2)}$ & 0.073 & 0.888 \\
Illinois Effect & $\frac{1}{7} \sum_{a=1}^{7} \hat{\theta}_{a+23,a}^{(2)}$ & 1.787 & 0.058 \\
\end{tabular}
\end{center}
\end{table}

The results of the analysis comport broadly with the results found in similar lottery studies, especially \cite{fuller_assessing_2022}, with modest but non-significant positive effects overall, especially in the first few weeks after full implementation. The state-specific results are also directionally consistent, with Illinois and Missouri having the largest positive estimates, Ohio a nearly zero estimated effect, and Michigan a small negative estimate. All states individually and considered together have larger estimates for the effect in the second week of the policy (denoted ``Second-Week Effect'' in Table~\ref{tbl:vax1}) compared to in the first adoption week or to the overall effect. This may indicate a strong treatment effect in the first full week of the policy which then wanes and zeroes out over time, consistent with the hypothesis of some authors (e.g., \cite{brehm_ohio_2022}) that the primary effect is to pull vaccinations that would have happened anyway earlier in time. The ability to examine state- and time-heterogeneity is thus particularly important for this research question.

We can additionally compare these results to those under existing staggered adoption methods. A two-way fixed-effects analysis gives a much larger estimated treatment effect of 1.703. This is hard to interpret, however, because of the composition across the time-varying treatment effects. The CS and SA methods all give estimates between 0.45 and 0.60, with permutation p-values between 0.25 and 0.55. These are broadly similar to many of the summary effects estimated using the generalized DID. The first-period effect estimators all estimate effects around 0.22, with p-values around 0.2. Again, this is similar to the first-period effect estimated by the generalized DID, although the latter has a lower p-value, which may reflect higher power.

The working covariance matrix chosen has a larger effect on the estimates here than in the SWT case. This is likely because of the strong auto-correlation of a cumulative outcome and the large heterogeneity between clusters. See Table~\ref{tbl:vaxS1} and Figure~\ref{fig:vaxWS1} in Appendix~\ref{appx:extrares} for estimates of the same estimands and heat maps of observation weights, respectively, under an independent working correlation structure. Note that the observation weights under AR(1) shown in Figure~\ref{fig:vaxW} have higher weight immediately prior to adoption, making them more similar to the CS and SA staggered adoption analysis methods. Additionally, the parallel trends assumption needs to be carefully considered. In this area, existing estimators may have advantages as they restrict comparison groups more strictly than does the generalized DID. So the investigator would need to weigh the trade-offs of potentially higher precision from incorporating more comparisons against the bias introduced by potential violations of parallel trends, preferably resulting in a pre-specified analysis plan.

\section{Simulation Results} \label{sec:sims}
To assess the operating characteristics of these estimators, I conduct a simulation study based on the SWT described in Section \ref{sec:SWTex} and the simulation study conducted in \cite{kennedyshaffer_novel_2020}. An SWT with 14 clusters and 8 time periods is simulated, where two clusters cross over to the intervention at each period starting in period 2. For cluster $i$ in period $j$, the outcomes for 100 individuals are generated according to the model:
\begin{align*}
\mu_{ij} &= \mu + \alpha_i + b_j + \nu_{ij} + \theta_{ij} X_{ij}, \\
Y_{ijk} &\overset{i.i.d.}{\sim} N(\mu_{ij}, \sigma^2_e),
\end{align*}
where $Y_{ijk}$ is the outcome of individual $k$ within cluster $i$ in period $j$, $X_{ij}$ is an indicator of treatment for cluster $i$ in period $j$, and $\nu_{ij} \overset{i.i.d}{\sim} N(0, \sigma^2_\nu)$. The following are used as fixed parameters: $\sigma^2_\nu = 0.01^2$, $\mu = 0.30$, $\bm{b} = \begin{pmatrix} 0 & 0.08 & 0.18 & 0.29 & 0.30 & 0.27 & 0.20 & 0.13 \end{pmatrix}$, and $\sigma^2_e = 0.1^2$. The random effects for cluster, $\alpha_i$, are sampled without replacement from:
\[ \bm{a} = \begin{pmatrix} -0.016, & -0.012, & -0.011, & -0.007, & -0.005, & -0.003, & -0.001, \\ 0, & 0.002, & 0.003, & 0.005, & 0.008, & 0.017, & 0.020 \end{pmatrix}  \]
These values have a mean of 0 and a standard deviation around 0.01. This generating process reflects a fixed set of clusters randomly assigned to intervention order, matching the randomization test approach used for inference.

The treatment effects $\theta_{ij}$ are varied by simulation scenario, with scenarios 1--3 having homogeneous treatment effects (akin to assumption S5), scenarios 4--6 having heterogeneous treatment effects by calendar time only (akin to assumption S4), and scenarios 7--9 having heterogeneous treatment effects by exposure time only (akin to assumption S3). Details are in Table \ref{tbl:scenarios}.
\begin{table}[!ht]
\caption{Treatment effect values used in the simulation scenarios. For calendar-time heterogeneous scenarios, the treatment effects are for periods starting with period 2; for exposure-time heterogeneous scenarios, the treatment effects are for the periods since intervention, starting with the first intervention period.\label{tbl:scenarios}}
\begin{center}
\begin{tabular}{r|ll}
Scenario & Heterogeneity & $\theta_{ij}$ Values \\
\hline
1 & Homogeneous & $\theta^{(5)} = 0$ \\
2 & Homogeneous & $\theta^{(5)} = -0.02$ \\
3 & Homogeneous & $\theta^{(5)} = -0.04$ \\
4 & Calendar-Time & $\theta^{(4)}_j = \begin{pmatrix} -0.07 & -0.05 & -0.03 & -0.01 & 0.01 & 0.03 & 0.05 \end{pmatrix}$ \\
5 & Calendar-Time & $\theta^{(4)}_j = \begin{pmatrix} -0.07 & -0.06 & -0.04 & 0 & 0.03 & 0.02 & 0.01 \end{pmatrix}$ \\
6 & Calendar-Time & $\theta^{(4)}_j = \begin{pmatrix} -0.03 & -0.03 & -0.03 & -0.03 & 0 & 0 & 0 \end{pmatrix}$ \\
7 & Exposure-Time & $\theta^{(3)}_a = \begin{pmatrix} -0.010 & -0.015 & -0.020 & -0.025 & -0.030 & -0.035 & -0.040 \end{pmatrix}$ \\
8 & Exposure-Time & $\theta^{(3)}_a = \begin{pmatrix} 0 & 0 & -0.03 & -0.03 & -0.03 & -0.03 & -0.03 \end{pmatrix}$ \\
9 & Exposure-Time & $\theta^{(3)}_a = \begin{pmatrix} -0.07 & -0.05 & -0.03 & -0.01 & 0.01 & 0.03 & 0.05 \end{pmatrix}$ \\
\hline
\end{tabular}
\end{center}
\end{table}

\begin{figure}[!ht]
\begin{center}
\includegraphics[width=7in]{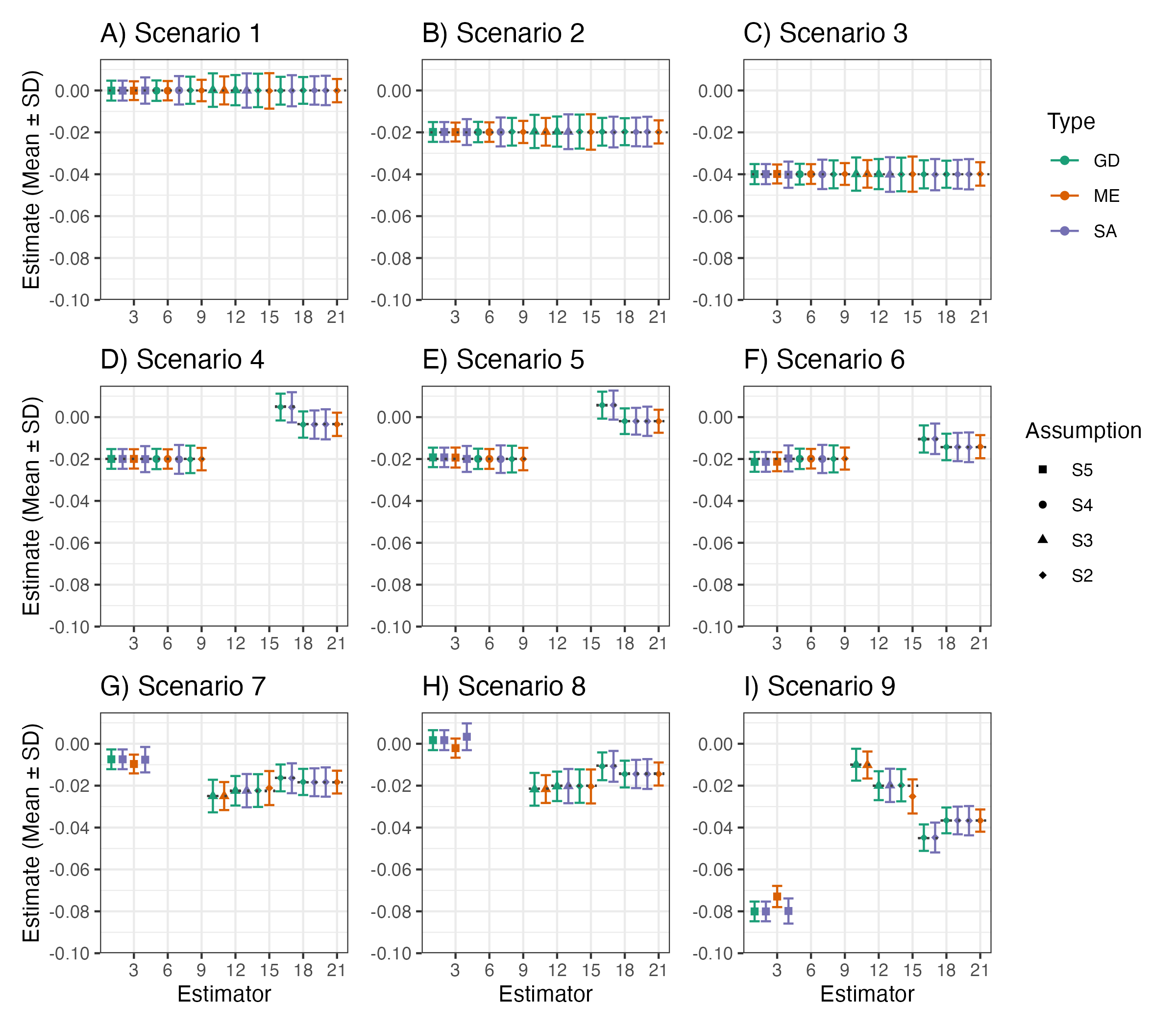}
\end{center}
\caption{Simulation results (mean estimate plus or minus one standard deviation of estimates across simulations) for 1,000 simulations each, by simulation scenario and estimator. Dotted horizontal lines represent the true value of the target estimand.\label{fig:sim_ests}}
\end{figure}

\begin{figure}[!ht]
\begin{center}
\includegraphics[width=7in]{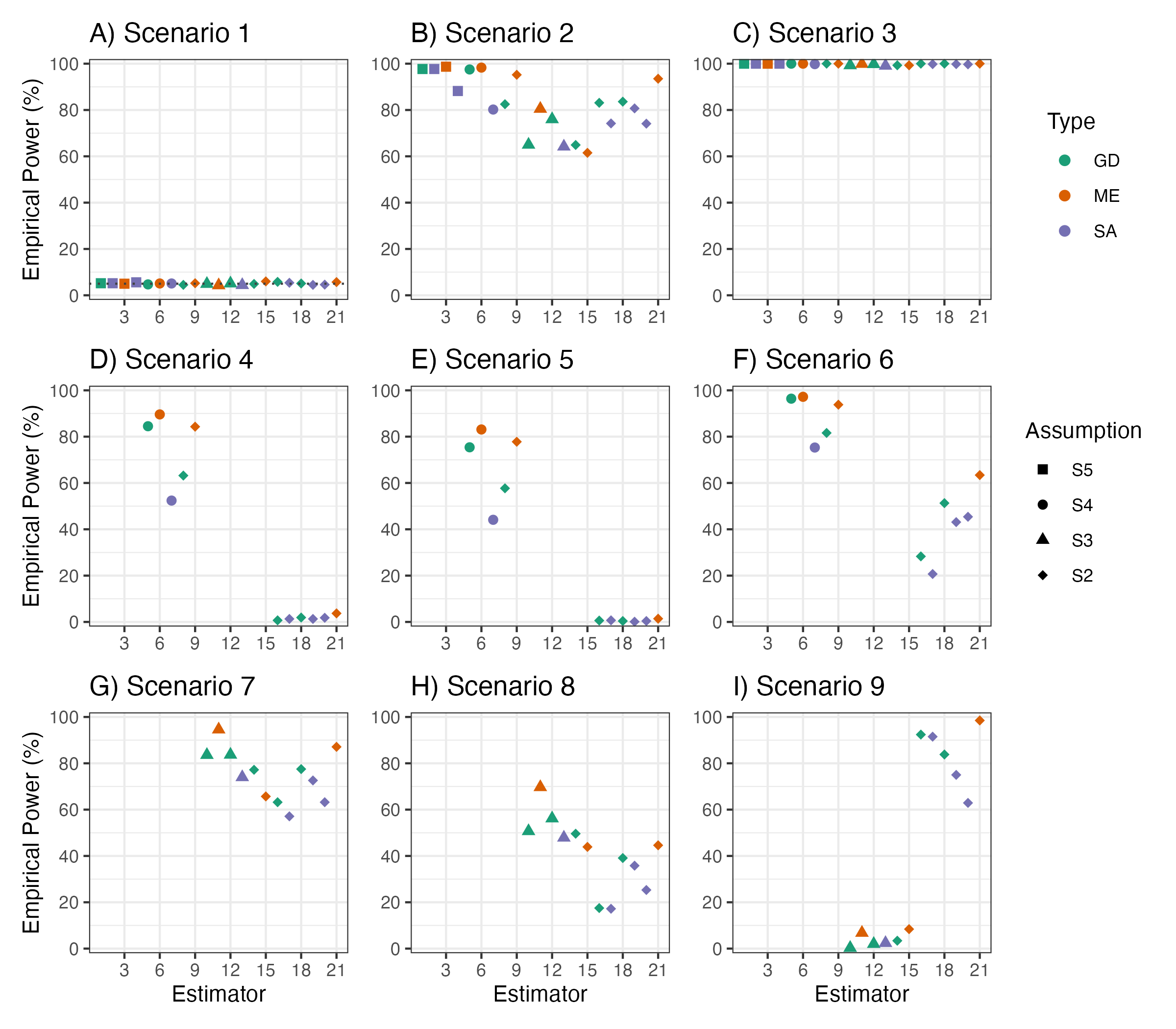}
\end{center}
\caption{Simulation results (empirical power at $\alpha=0.05$ level) for 1,000 simulations each, by simulation scenario and estimator. Note that biased estimators or those targeting substantially different estimands are excluded.\label{fig:sim_power}}
\end{figure}

For each scenario, estimators using the correct assumption setting for that scenario are considered, as well as those with more flexible assumptions to determine the effect on efficiency of more robust methods. The estimators are categorized by their assumption setting and type (``GD'' represents a generalized DID estimator, ``CPI'' an estimator using a mixed effects model with a cluster-time random effect, and ``SA'' an estimator from the existing staggered adoption literature). All generalized DID estimators were calculated assuming independent observations; results for an exchangeable covariance structure were very similar. Full details on the estimators used are in Appendix~\ref{appx:simDets}. For Scenarios 1--3, all estimators are included. Scenarios 4--9 exclude estimators targeting the other heterogeneity only. All scenarios were simulated 1000 times. Mean estimates plus or minus one standard deviation are plotted in Figure~\ref{fig:sim_ests}. For each simulation, a randomization-based hypothesis test was conducted with 250 permutations; those with p-values less than $\alpha = 0.05$ were considered statistically significant, and the empirical power results (or Type I Error for null effects) for each are plotted in Figure~\ref{fig:sim_power}. Appendix~\ref{appx:extrares} includes results for estimators targeting specific period effects.

In general, across simulations, most estimators were unbiased for the target estimand. When assuming full treatment homogeneity (Assumption S5), however, the estimators do not target an estimand within the convex hull of treatment effects when treatment effect varies by exposure time (so-called ``dynamic treatment effects'', Scenarios 7--9). Estimators using the generalized DID method, two-way fixed effects (\cite{goodman-bacon_difference--differences_2021}), or the crossover method CO-3 (\cite{kennedyshaffer_novel_2020}) target a linear combination of treatment effects that can be identified (see Estimators 1, 2, and 4, respectively, in Figs.~\ref{fig:sim_ests} and \ref{fig:sim_power}); the mixed effects model estimator appears to target a different estimand that is not currently identified. In addition, with dynamic treatment effects, a mixed effects model fully saturated with calendar time- and exposure time-heterogeneous treatment effects averaged across exposure times (Estimator 15 in Figs.~\ref{fig:sim_ests} and \ref{fig:sim_power}) is not unbiased for the exposure-time-averaged treatment effect.

Aside from those cases, general patterns appear for the estimators. In general, for a certain estimand and assumption setting, existing staggered adoption methods are less efficient and less powerful than generalized DID estimators, which, in turn, are less efficient and less powerful than mixed effects model estimators. For targeting specific period effects or averages, there are cases where the generalized DID estimator can be more efficient than the corresponding mixed effects model estimator; e.g., estimators 14 and 15 in Figs.~\ref{fig:sim_ests} and \ref{fig:sim_power} and estimators 7, 8, 12, and 13 in Figs.~\ref{fig:sim_t_ests} and \ref{fig:sim_t_power}. Note that for all settings considered here, the mixed effects model is (nearly) correctly specified in both its fixed effect and random effect terms. If the random effect term is misspecified to a greater extent, there may be either bias or, more commonly, reduced efficiency of these estimators.

\section{Discussion} \label{sec:disc}

The proposed method provides a highly flexible and adaptable approach to analyzing stepped wedge and staggered adoption settings. In particular, it allows the investigator to specify a target estimand that is any linear combination of unique treatment effects and guarantees unbiasedness under the standard causal assumptions and the particular treatment effect heterogeneity assumptions specified. Among this class of unbiased estimators, the investigator can then identify the one with the lowest variance for an assumed working covariance structure. Analogous to the working correlation matrix used in generalized estimating equations (\cite{zeger_longitudinal_1986}), this working covariance structure need not be correctly specified to maintain unbiased estimation. A misspecified structure, however, may lead to suboptimal variance and reduced power to detect an effect. This framework unifies ideas from both the observational staggered adoption methods and randomized SWT methods, providing estimators with more flexibility and interpretability under heterogeneity than SWT estimators and lower variance than staggered adoption methods.

The composition of the estimator from two-by-two DID estimators serves several purposes. First, it handles unit- and time-confounding and ensures the use of both horizontal and vertical comparisons, as discussed by \cite{matthews_stepped_2017}. Second, it allows both intuitive connections to methods commonly used in econometrics and quantitative social science for quasi-experimental panel data as well as assumptions for validity formulated similarly to those in the existing literature of that field. Third, it allows clear formulation of the treatment effect heterogeneity assumptions that are made rather than those assumptions arising implicitly through a linear regression formulation. In fact, these building blocks, including non-canonical two-by-two comparisons, have independently been proposed as useful methods to test for heterogeneity of treatment effects (\cite{shahn_group_2024,xu_factorial_2024}). By specifying the variance assumption separately, it also removes the need for a correct joint specification of the mean and variance models and allows those assumptions (for the working structure) to be formulated independently of the treatment effect assumptions. These approaches also allow for ease of sensitivity analysis, as the expectation can be found using Equation~\ref{eqn:e-theta-hat} using a specified weight vector under any other assumption setting. In particular, unlike the mixed effects models that are commonly used for SWT analysis, the precise estimand targeted if there are additional unspecified heterogeneities can be identified, allowing more robust interpretation. However, this comes at a small cost to power and precision compared to a correctly-specified mixed effects model, as shown in the simulations. Along with the potential to assess relative efficiency under the working covariance structure as shown in the examples, these properties allow the bias-variance tradeoff to be made explicit, and permit investigators to consider the appropriate generalizability and interpretation of their target estimands.

The example re-analyzing the data collected by \cite{trajman_impact_2015} demonstrates the use of this method for a variety of target estimands in a SWT. Results largely align with those of existing non-parametric estimators, which already had been shown to have desirable bias and inference properties (\cite{thompson_robust_2018, kennedyshaffer_novel_2020}). In nearly comparable situations, like the assumption of a homogeneous effect compared to the inverse-variance weighted average vertical approach (\cite{thompson_robust_2018}) or either crossover-type estimator (\cite{kennedyshaffer_novel_2020}), this approach had a lower p-value, which may reflect improved precision, as shown in the simulations. In addition, many other estimates are tractable using this approach, allowing for flexible pre-specified or exploratory analysis of treatment effect heterogeneity.

The observational staggered adoption setting using data collected and analyzed by \cite{fuller_assessing_2022} demonstrates that this method can also be used for non-randomized settings. Results again directionally align with those identified using various synthetic control and DID estimators, but this method allows more specific targeting of certain effects while also making clear the homogeneity assumptions being made. These are largely similar to the results obtained using recently-proposed staggered adoption methods by \cite{callaway_difference--differences_2021} and \cite{sun_estimating_2021}, but allow for the use of more data by not restricting to the last-untreated period. The simulations demonstrate that, if the assumptions are met, the generalized DID estimator tends to be more efficient than corresponding estimators of the forms described by \cite{callaway_difference--differences_2021} and \cite{sun_estimating_2021}. Again, it is important to consider the parallel trends assumption here as well in order to interpret the estimates causally. Trade-offs between efficiency and validity arise due to these additional considerations in the non-randomized case, and further exploration of this specific consideration is warranted.

The simulations demonstrate the feasibility of using these estimators in the analysis of stepped wedge trials and the tradeoffs compared to existing mixed effects model approaches and existing staggered adoption approaches. For non-randomized settings, simulation under a reasonable data-generating process should be conducted to confirm these general results. The improved efficiency compared to purely vertical, pure crossover approaches, or staggered adoption approaches that limit the comparisons used, however, should all remain, as this approach uses as much information as possible under the assumed treatment effect heterogeneities. The building blocks enforce zero-sums on weights in both rows and columns, as discussed by \cite{matthews_stepped_2017}. In addition, the fact that the method generally highly weights cells with high information content (\cite{kasza_information_2019, kasza_information_2019-1}) and respects centrosymmetry under appropriate assumptions (\cite{bowden_centrosymmetry_2021}) indicates efficient use of the observed data. However, their performance compared to regression-based approaches (e.g., \cite{hooper_sample_2016,lindner_heterogeneous_2021,kenny_analysis_2022,maleyeff_assessing_2023,lee_cluster_2024}) and existing robust estimators (e.g., \cite{hughes_robust_2020,roth_efficient_2023,borusyak_revisiting_2024}) remains to be determined. In particular, future study and setting-specific simulations are needed to explore when this approach provides equivalent estimators to those approaches and when one outperforms the other. Further work is also needed in the non-randomized case to determine the differences in efficiency and in the assumptions required between these approaches and existing staggered adoption estimators, extending the work in Section~\ref{sec:Comps} to a variety of settings.

Additional future work should consider identifying other approaches to inference, including the possibility of closed-form variance estimators and asymptotic sampling distributions under certain variance assumptions. Identifying appropriate plug-in estimators for use in Equation~\ref{eqn:var-plugin}, their robustness to misspecification of the working covariance structure, and the asymptotic distribution would be key, as has been done for mixed effects models with misspecification (see, e.g., \cite{voldal_model_2022}). This would allow for more targeted design of experiments when analysis will proceed using these methods. Understanding likely covariance structures, especially given planned or conducted approaches for sampling of units (see, e.g., \cite{hooper_key_2021}), would improve selection of $\bm{M}$ as well. This would also enable either simulation or analytic approaches to sample size determination, a key feature in study design. Additionally, modifications of this method to accommodate adjustment for covariates would be useful for non-randomized staggered adoption settings where parallel trends holds only conditionally (see, e.g., \cite{callaway_difference--differences_2021,roth_whats_2023}) and SWTs with restricted, stratified, or matched randomization (see, e.g., \cite{copas_designing_2015}).

Both stepped wedge and staggered adoption settings are common study designs with important roles to play in assessing a variety of policies and implementation approaches; for example, both have been proposed as useful tools for rapid policy evaluation in the COVID-19 pandemic or similar settings (see, e.g., \cite{goodman-bacon_using_2020,bell_equity_2021,cowger_lifting_2022,kennedy-shaffer_quasi-experimental_2024}). They each, however, have statistical intricacies that should be carefully considered in the design and analysis (see, e.g., \cite{baker_how_2022,nevins_adherence_2024}). The generalized DID estimator framework proposed here has desirable bias and variance properties and interpretability that enable its use across many such settings and desired treatment effect estimands. 

\subsection*{Data Availability Statement}
All code used for this manuscript is available at \url{https://github.com/leekshaffer/GenDID}. Included there is a simulated data set based on the stepped-wedge trial data example. Access policies for the original data set can be found in \cite{trajman_impact_2015} at \url{https://doi.org/10.1371/journal.pone.0123252}. The original data set for the staggered adoption example are included at \url{https://github.com/leekshaffer/GenDID}; details on the collection of that data can be found in \cite{fuller_assessing_2022} at \url{https://doi.org/10.1371/journal.pone.0274374}. The original data repository for that article is available at \url{https://doi.org/10.7910/DVN/K1XX02}; the file used is \verb|synth_data_clean.Rdata|. See the repository for citation, access, and sharing policies.

\subsection*{Acknowledgements}
The author wishes to thank Prof.\ Anete Trajman for making the stepped-wedge trial data available to the author for re-analysis and Prof.\ Jennifer Thompson for helpful discussions regarding the data. The author wishes to thank Sam Fuller, Sara Kazemian, Prof.\ Carlos Algara, and Prof.\ Daniel J. Simmons for making their COVID-19 vaccine lottery data publicly available. The author also wishes to thank colleagues at Vassar College (specifically, Profs.\ Andy Borum, Adam Lowrance, Lisa Lowrance, and Benjamin Morin and students Raymond Aye, Ryan Kilpadi, and Andrew Steindl) for helpful discussions on linear algebra and colleagues at the Yale School of Public Health (specifically, Profs.\ Fan Li and Donna Spiegelman) for helpful discussions on stepped-wedge trial settings. Finally, the author wishes to thank participants in the Current Developments in Cluster Randomised Trials and Stepped Wedge Designs conference (Queen Mary University of London, November 2022), the Fourth International Conference on Stepped Wedge Design (University of York, May 2024), and the University of Illinois Department of Statistics seminar series (September 2024) for helpful feedback on this project.

\clearpage

\bibliographystyle{agsm}
\bibliography{main}

\clearpage

\bigskip
\begin{center}
{\large\bf APPENDICES}
\end{center}

\appendix

\section{Determining the Matrix of DID Estimators} \label{appx:A_const}
For any integer $j \ge 2$, define $\bm{A}_j$ as the $j-1 \times j$ matrix where:
\[ \bm{A}_j = \begin{pmatrix} -1 & 1 & 0 & 0 & \cdots & 0 \\ -1 & 0 & 1 & 0 & \cdots & 0 \\ \vdots & \vdots & & \ddots & \vdots \\ -1 & 0 & 0 & 0 & \cdots & 1 \end{pmatrix} = \begin{pmatrix} - \bm{1}_{j-1} & | & \bm{I}_{j-1}, \end{pmatrix}\]
where $\bm{1}_{k}$ is the $k \times 1 $ column vector of 1's and $\bm{I}_{k}$ is the $k \times k$ identity matrix.

Now define $\bm{A}_\cdot$ for a setting with $J$ time periods as the $\binom{J}{2} \times J$ matrix given by:
\[ \bm{A}_\cdot = \begin{pmatrix} \multicolumn{3}{c}{\bm{A}_J} \\ \bm{0}_{J-2} & \multicolumn{2}{c}{\bm{A}_{J-1}} \\ \bm{0}_{J-3} & \bm{0}_{J-3} & \bm{A}_{J-2} \\ & \vdots & \\ \bm{0}_1 & \cdots & \bm{A}_2\end{pmatrix},\]
where $\bm{0}_k$ is the $k \times 1$ column vector of 0's.

Finally, define $\bm{A}$ for a setting with $J$ time periods and $N$ clusters as the $\binom{N}{2} \binom{J}{2} \times NJ$ matrix given by:
\[ \bm{A} = \begin{pmatrix} \multicolumn{2}{c}{\bm{A}_\cdot} & \multicolumn{2}{c}{-\bm{A}_\cdot} & \bm{0}_{\binom{J}{2} \times (N-2) J} \\
\multicolumn{2}{c}{\bm{A}_\cdot} & \bm{0}_{\binom{J}{2} \times J} & -\bm{A}_\cdot & \bm{0}_{\binom{J}{2} \times (N-3) J} \\
\multicolumn{2}{c}{\bm{A}_\cdot} & \bm{0}_{\binom{J}{2} \times 2 J} & -\bm{A}_\cdot & \bm{0}_{\binom{J}{2} \times (N-4) J} \\
& & \vdots & & \\
\multicolumn{2}{c}{\bm{A}_\cdot} & \bm{0}_{\binom{J}{2} \times (N-2)J} & \multicolumn{2}{c}{-\bm{A}_\cdot} \\
\bm{0}_{\binom{J}{2} \times J} & \bm{A}_\cdot & \multicolumn{2}{c}{-\bm{A}_\cdot} & \bm{0}_{\binom{J}{2} \times (N-3) J} \\
\bm{0}_{\binom{J}{2} \times J} & \bm{A}_\cdot & \bm{0}_{\binom{J}{2} \times J} & -A_\cdot & \bm{0}_{\binom{J}{2} \times (N-4) J} \\
& & \vdots & & \\
\bm{0}_{\binom{J}{2} \times (N-2) J} & \multicolumn{2}{c}{\bm{A}_\cdot} & \multicolumn{2}{c}{-\bm{A}_\cdot}
\end{pmatrix}, \]
where $\bm{0}_{k \times k'}$ is the $k \times k'$ matrix of 0's. Note that, in the above representation, each row represents $\binom{J}{2}$ rows. In each of these, $(N-2) J$ columns are from $\bm{0}$ matrices, while the other $2J$ are from two copies of the $\bm{A}_\cdot$ matrix.

This is the matrix that, when multiplied (on the right) by the column vector of outcomes $\bm{Y}$, gives the column vector of two-by-two DID estimators $\bm{d}$, defined in Section~\ref{sec:meth-OEF}. In particular, $D_{i,i',j,j'}$ will be found on the following row number of the vector $\bm{d}$:
\begin{align*}
    1 + \binom{J}{2} (i-1) \left( N-\frac{i}{2} \right) + (i' - i - 1) \binom{J}{2} + (j-1) \left( J-\frac{j}{2} \right) + (j' - j -1)
\end{align*}

Conversely, row $k$ in the vector $\bm{d}$, for $1 \le k \le \binom{J}{2} \binom{N}{2}$, corresponds to the two-by-two DID estimator $D_{i,i',j,j'}$ where the indices are given by the following algorithm:
\begin{enumerate}
\item Let $c_1 = \floor{\frac{k}{\binom{J}{2}}} + 1$ and $c_2 = k \mbox{ mod } \binom{J}{2} + 1$
\item Let $n^*$ be the minimum $n \in \{1,2,\ldots,N\}$ such that $c_1 \le \sum_{\ell=1}^n (N-\ell)$. Then $i=n^*$ and $i' = i + c_1 - \sum_{\ell=1}^{n^*-1} (N-\ell)$.
\item Let $m^*$ be the minimum $m \in \{1,2,\ldots,J\}$ such that $c_2 \le \sum_{\ell=1}^m (J-\ell)$. Then $j=m^*$ and $j' = j + c_1 - \sum_{\ell=1}^{m^*-1} (J-\ell)$.
\end{enumerate}

Each row $k$ can then be associated with the corresponding $D_{i,i',j,j'}$. From the values of $T_i$ and $T_{i'}$, along with $j$ and $j'$, the type (from 1--6) of this two-by-two DID estimator can then be determined using Table~\ref{tbl:2x2-cats}. And its expectation under the specified assumption setting can be determined as well using Table~\ref{tbl:exp2}. Thus, the $\binom{J}{2} \binom{N}{2} \times 1$ column vector $E[\bm{d}]$ and associated matrix $\bm{F}$ can be constructed in terms of the estimands possible for the assumption setting.

\clearpage

\section{Proofs of Key Lemmata and Theorems} \label{appx:Proofs}

\subsection{Lemmata for Proof of Theorem~\ref{thm:rank}}

\textbf{Lemma \ref{thm:rank}.1} Let $\bm{F}$ and $\bm{A}$ be as defined in Section~\ref{sec:meth}. Then $ker(\bm{A}^T) \subset ker(\bm{F}^T)$.

\noindent \emph{Proof.} Let $\bm{x} \in ker(\bm{A}^T)$; that is, $\bm{A}^T \bm{x} = \bm{0}$. Then $\bm{d}^T \bm{x} = \left( \bm{A} \bm{y} \right)^T \bm{x} = \bm{y}^T \bm{A}^T \bm{x} = \bm{y}^T \bm{0} = 0$ for any $\bm{y}$. Since this is true for any $\bm{y}$, it must be true in expectation, so $0 = E[\bm{d}^T \bm{x}] = E[\bm{d}^T] \bm{x} = E[\bm{d}]^T \bm{x} = \left( \bm{F} \bm{\theta} \right)^T \bm{x} = \bm{\theta}^T \bm{F}^T \bm{x}$. For this to be true for any vector of treatment effects, then, $\bm{F}^T \bm{x} = \bm{0}$ and $\bm{x} \in ker(\bm{F}^T)$. So $ker(\bm{A}^T) \subset ker(\bm{F}^T)$, as desired.

\noindent \textbf{Lemma \ref{thm:rank}.2} Let $\bm{A}$ be as defined in Section~\ref{sec:meth} for a setting with $N \ge 2$ units and $J \ge 2$ periods. Then $rank(\bm{A}) = rank(\bm{A}^T) = (N-1) (J-1)$.

\noindent \emph{Proof.} For any $i < i'$, $j < j'$, let $\bm{A}_{i,i',j,j'}$ be the row of $\bm{A}$ corresponding to $D_{i,i',j,j'}$ (i.e., $\bm{A}_{i,i',j,j'}^T \bm{y} = D_{i,i',j,j'}$). $\bm{A}_{i,i',j,j'}$ has the value 1 in the columns corresponding to $Y_{ij'}$ and $Y_{i'j}$, -1 in the columns corresponding to $Y_{ij}$ and $Y_{i'j'}$, and 0 in all other columns.

Consider the matrix $\bm{A}^*$ composed of only the rows of the form $\bm{A}_{1,i',1,j'}$ of $\bm{A}$, where $2 \le i' \le N$ and $2 \le j' \le J$. This is a $(N-1)(J-1) \times NJ$ matrix. Any row $\bm{A}_{i,i',j,j'}$ can be expressed as a linear combination of the rows of $\bm{A^*}$ as follows:
\[ \bm{A}_{i,i',j,j'} = \left( \bm{A}_{1,i',1,j'} - \bm{A}_{1,i,1,j'} \right) - \left( \bm{A}_{1,i',1,j} - \bm{A}_{1,i,1,j} \right). \]
Thus, $rank(\bm{A}) = rank(\bm{A}^*)$. Moreover, the rows of $\bm{A^*}$ are linearly independent, since each row of the form $\bm{A}_{1,i',1,j'}$ is the unique row of that form to have a non-zero entry in the column corresponding to $Y_{i'j'}$. So $rank(\bm{A}) = rank(\bm{A^*}) = (N-1)(J-1)$. Since $\bm{A}$ is a real matrix, $rank(\bm{A}^T) = rank(\bm{A}) = (N-1)(J-1)$ as well.

\clearpage

\subsection{Proof of Theorem~\ref{thm:rank}}

\noindent \textbf{Theorem~\ref{thm:rank}} Let $\bm{F}$, $\bm{w}$, $\bm{v}$, and $\bm{d}$ be as defined in Section~\ref{sec:meth}. Then the following are true about the set of estimators of the form $\hat{\theta} = \bm{w}^T \bm{d}$ that are unbiased for $\theta_e$ under the assumption setting:
\begin{itemize}
\item If $rank(\bm{F}^T|\bm{v}) > rank(\bm{F}^T)$, then there are no estimators $\hat{\theta}$ of this form that are unbiased for $\theta_e$.
\item If $rank(\bm{F}^T|\bm{v}) = rank(\bm{F}^T) = \binom{N}{2} \binom{J}{2}$, then there is a unique estimator $\hat{\theta}$ of this form that is unbiased for $\theta_e$, defined by the unique $\bm{w}$ that solves $\bm{F}^T \bm{w} = \bm{v}$.
\item If $rank(\bm{F}^T|\bm{v}) = rank(\bm{F}^T) < \binom{N}{2} \binom{J}{2}$, then there are infinitely many estimators $\hat{\theta}$ of this form that are unbiased for $\theta_e$. The dimension of unique such estimators is $(N-1)(J-1) - rank(\bm{F})$.
\end{itemize}

\noindent \emph{Proof.} By results on non-homogeneous systems of linear equations (also known as the Rouch\'e-Capelli Theorem; see \cite{george_course_2024}, p. 41) on the equation $\bm{F}^T \bm{w} = \bm{v}$, there are two possibilities:
\begin{itemize}
\item If $rank(\bm{F}^T|\bm{v}) > rank(\bm{F}^T)$, then there are no solutions.
\item If $rank(\bm{F}^T|\bm{v} = rank(\bm{F}^T)$, then there is a solution to the system, and the affine space of solutions $W$ has dimension $\binom{N}{2} \binom{J}{2} - rank(\bm{F}^T)$, since $\bm{F}^T$ is of dimension $\lvert \bm{\theta} \rvert \times \binom{N}{2} \binom{J}{2}$, where $\lvert \bm{\theta} \rvert$ is the number of unique treatment effects.
\end{itemize}

If $rank(\bm{F}^T) = \binom{N}{2} \binom{J}{2}$, then, there is a single unique solution $\bm{w}$, which corresponds to a single unique estimator $\hat{\theta}$ of the desired form that is unbiased for $\theta_e$.

If $rank(\bm{F}^T) < \binom{N}{2} \binom{J}{2}$, there are infinitely many solutions $\bm{w}$. Two distinct solutions $\bm{w}_1 \neq \bm{w}_2$, however, may correspond to the same estimator $\hat{\theta}$, since different weightings of the two-by-two DID estimators may result in the same weightings of the underlying observations.

Let $\bm{w}_1 \in W$ be a solution to $\bm{F}^T \bm{w} = \bm{v}$. Any other solution to $\bm{F}^T \bm{w} = \bm{v}$ can be expressed as $\bm{w}_1 + \bm{x}$, where $\bm{x} \in ker(\bm{F}^T)$. The vector space of $ker(\bm{F}^T)$ has dimension $nullity(\bm{F}^T) = \binom{N}{2} \binom{J}{2} - rank(\bm{F}^T)$, corresponding to the dimension of the affine space $W$ as found above. By Lemma~\ref{thm:rank}.1, $ker(\bm{A}^T) \subset ker(\bm{A}^T)$, and so $ker(\bm{F}^T)$ can be written as the direct sum of $ker(\bm{A}^T) + K$, where $K$ is defined as the orthogonal complement of $ker(\bm{A}^T)$ within $ker(\bm{F}^T)$. So we can further express $W = \{\bm{w}_1+\bm{x}_1+\bm{x}_2:~\bm{x}_1 \in ker(\bm{A}^T),~\bm{x}_2 \in K\}$. However, $\bm{A}^T \bm{x}_1 = \bm{0}$ if $\bm{x}_1 \in ker(\bm{A}^T)$. Hence, for any $\bm{w} \in W$ and any $\bm{x}_1 \in ker(\bm{A}^T)$, $\bm{w}^T \bm{A} = (\bm{w} + \bm{x}_1)^T \bm{A}$, and so the estimators $\hat{\theta}_1$ and $\hat{\theta}_2$ defined by the weight vectors $\bm{w}$ and $\bm{w} + \bm{x}_1$ are equal for all $\bm{y}$. Thus, the only unique estimators (in terms of the underlying observations) are given by the subspace $W_A = \{\bm{w}_1 + \bm{x}_2:~\bm{x}_w \in K\}$.

Thus, the affine space of \textit{unique} estimators $\hat{\theta}$ that are unbiased for the desired $\theta_e$ under the specified assumption setting has dimension given by:
\[ dim(K) = dim(W) - dim(ker(\bm{A}^T)) = nullity(\bm{F}^T) - nullity(\bm{A}^T). \]
This can be further simplified using the rank-nullity theorem (see \cite{george_course_2024}, p. 89) and Lemma~\ref{thm:rank}.2:
\begin{align*}
dim(K) &= dim(W) - dim(ker(\bm{A}^T)) = nullity(\bm{F}^T) - nullity(\bm{A}^T) \\
&= \left( \binom{N}{2} \binom{J}{2} - rank(\bm{F}^T) \right) - \left( \binom{N}{2} \binom{J}{2} - rank(\bm{A}^T) \right) \\
&= rank(\bm{A}^T) - rank(\bm{F}^T) = (N-1)(J-1) - rank(\bm{F}),
\end{align*}
as desired.

\clearpage

\section{Details of Simulation Estimators} \label{appx:simDets}

The overall effect estimators used in the simulations are detailed in Table~\ref{tbl:SimOverallEsts}.

\begin{table}[!ht]
\caption{Estimators targeting overall effects used for comparisons in simulations. Number corresponds to the x-axis in Figs.~\ref{fig:sim_ests} and \ref{fig:sim_power}. Type corresponds to the labels in those figures, with GD representing a generalized DID estimator, SA a staggered adoption estimator (see Section~\ref{sec:Comps}), and ME a mixed effects model estimator. Assumption corresponds to the effect heterogeneity assumption setting outlined in Table~\ref{tbl:assnset}; for non-GD estimators, the assumption setting is inferred from the model and estimator. Unless otherwise noted: $a_i$ and $b_j$ refer to fixed effects for cluster $i$ and period $j$, respectively; $\alpha_i$, $\nu_{ij}$, and $\epsilon_{ijk}$ refer to random effects for cluster, cluster-period, and individual respectively and follow independent normal distributions with mean 0 and variances $\sigma^2_\alpha$, $\sigma^2_\nu$, and $\sigma^2_e$, respectively; and $X_{ij}$ is the binary treatment indicator for cluster $i$ in period $j$ with $T_i$ equal to the first period for cluster $i$ where $X_{ij} = 1$. Estimator for GD estimators uses the notation from this paper; for ME estimators, it uses the defined model; and for other estimators, it follows the cited source as closely as possible. \label{tbl:SimOverallEsts}}
\end{table}

\begin{center}
\begin{tabular}{r|lcp{60mm}|c|}
No. & Type & Assumption & Model & Estimator \\
\hline
 & \multicolumn{4}{c|}{Overall Estimand} \\
\hline
1 & GD & S5 & Independent Covariance & $\hat{\theta}^{(5)}$ \\
2 & SA & S5 & TW: two-way fixed effects\newline $E[Y_{ijk}] = a_i + b_j + \theta X_{ij}$ & $\hat{\theta}$ \\
3 & ME & S5 & $Y_{ijk} = \mu + \alpha_i + b_j + \nu_{ij} + \theta X_{ij} + \epsilon_{ijk}$ & $\hat{\theta}$\\
4 & SA & S5 & CO-3: equal weights \newline (\cite{kennedyshaffer_novel_2020}) & $\hat{\theta}^{CO-3}$\\
\hline
& \multicolumn{4}{c|}{Calendar-Time Averaged Estimand} \\
\hline
5 & GD & S4 & Independent Covariance & $\frac{1}{6} \sum_{j=2}^7 \hat{\theta}^{(4)}_j$ \\
6 & ME & S4 & $Y_{ijk} = \mu + \alpha_i + b_j + \nu_{ij} + \theta_j X_{ij} + \epsilon_{ijk}$ & $\frac{1}{6} \sum_{j=2}^7 \hat{\theta}_j$ \\
7 & SA & S4 & CS: calendar ATT \newline (\cite{callaway_difference--differences_2021}) & $\frac{1}{6} \sum_{t=2}^7 \hat{\theta}_c(t)$ \\
8 & GD & S2 & Independent Covariance & $\frac{1}{6} \sum_{j=2}^7 \left[ \frac{1}{j-1} \sum_{a=1}^{j-1} \hat{\theta}^{(2)}_{j,a} \right]$ \\
9 & ME & S2 & $Y_{ijk} = \mu + \alpha_i + b_j + \nu_{ij} + \theta_{j,a} X_{ij} I(T_i - j + 1 = a) + \epsilon_{ijk}$ & $\frac{1}{6} \sum_{j=2}^7 \left[ \frac{1}{j-1} \sum_{a=1}^{j-1} \hat{\theta}_{j,a} \right]$ \\
\hline
\end{tabular}

\begin{tabular}{r|lcp{60mm}|c|}
No. & Type & Assumption & Model & Estimator \\
\hline
& \multicolumn{4}{c|}{Exposure-Time Averaged Estimand} \\
\hline
10 & GD & S3 & Independent Covariance & $\frac{1}{7} \sum_{a=1}^7 \hat{\theta}^{(3)}_a$ \\
11 & ME & S3 & $Y_{ijk} = \mu + \alpha_i + b_j + \nu_{ij} + \theta_a X_{ij} I(T_i - j + 1 = a) + \epsilon_{ijk}$ & $\frac{1}{7} \sum_{a=1}^7 \hat{\theta}_a$ \\
12 & GD & S3 & Independent Covariance & $\frac{1}{6} \sum_{a=1}^6 \hat{\theta}^{(3)}_a$ \\
13 & SA & S3 & CS: dynamic ATT \newline (\cite{callaway_difference--differences_2021}) & $\frac{1}{6} \sum_{e=1}^6 \hat{\theta}_{es}(e)$ \\
14 & GD & S2 & Independent Covariance & $\frac{1}{6} \sum_{a=1}^6 \left[ \frac{1}{6-a} \sum_{j=a+1}^7 \hat{\theta}^{(2)}_{j,a} \right]$ \\
15 & ME & S2 & $Y_{ijk} = \mu + \alpha_i + b_j + \nu_{ij} + \theta_{j,a} X_{ij} I(T_i - j + 1 = a) + \epsilon_{ijk}$ & $\frac{1}{6} \sum_{a=1}^6 \left[ \frac{1}{6-a} \sum_{j=a+1}^{7} \hat{\theta}_{j,a} \right]$ \\
\hline
& \multicolumn{4}{c|}{Group Averaged Estimand} \\
\hline
16 & GD & S2 & Independent Covariance & $\frac{1}{6} \sum_{a=1}^6 \sum_{j=a+1}^7 \frac{1}{7-j+a} \hat{\theta}_{j,a}^{(2)}$ \\
17 & SA & S2 & CS: group ATT \newline (\cite{callaway_difference--differences_2021}) & $\hat{\theta}^O_{sel} = \frac{1}{6} \sum_{\tilde{g}=2}^7 \hat{\theta}_{sel}(\tilde{g})$ \\
\hline
& \multicolumn{4}{c|}{Average Treatment Effect on the Treated (Cluster-Period Averaged) Estimand} \\
\hline
18 & GD & S2 & Independent Covariance & $\frac{1}{21} \sum_{j=2}^7 \sum_{a=1}^{j-1} \hat{\theta}_{j,a}^{(2)}$ \\
19 & SA & S2 & CS: simple ATT \newline (\cite{callaway_difference--differences_2021}) & $\hat{\theta}^O_W$ \\
20 & SA & S2 & SA: Interaction-Weighted ATT \newline (\cite{sun_estimating_2021}) & $\frac{1}{6} \sum_{g=2}^7 \hat{v}_g$ \\
21 & ME & S2 & $Y_{ijk} = \mu + \alpha_i + b_j + \nu_{ij} + \theta_{j,a} X_{ij} I(T_i - j + 1 = a) + \epsilon_{ijk}$ & $\frac{1}{21} \sum_{a=1}^6 \sum_{j=a+1}^{7} \hat{\theta}_{j,a}$ \\
\hline
\end{tabular}
\end{center}

\clearpage

The period-specific effect estimators used in the simulations are detailed in Table~\ref{tbl:SimTargetEsts}.

\begin{table}[!ht]
\caption{Estimators targeting period-specific effects used for comparisons in simulations. Number corresponds to the x-axis in Figs.~\ref{fig:sim_t_ests} and \ref{fig:sim_t_power}. Type corresponds to the labels in those figures, with GD representing a generalized DID estimator, SA a staggered adoption estimator (see Section~\ref{sec:Comps}), and ME a mixed effects model estimator. Assumption corresponds to the effect heterogeneity assumption setting outlined in Table~\ref{tbl:assnset}; for non-GD estimators, the assumption setting is inferred from the model and estimator. Unless otherwise noted: $a_i$ and $b_j$ refer to fixed effects for cluster $i$ and period $j$, respectively; $\alpha_i$, $\nu_{ij}$, and $\epsilon_{ijk}$ refer to random effects for cluster, cluster-period, and individual respectively and follow independent normal distributions with mean 0 and variances $\sigma^2_\alpha$, $\sigma^2_\nu$, and $\sigma^2_e$, respectively; and $X_{ij}$ is the binary treatment indicator for cluster $i$ in period $j$ with $T_i$ equal to the first period for cluster $i$ where $X_{ij} = 1$. Estimator for GD estimators uses the notation from this paper; for ME estimators, it uses the defined model; and for other estimators, it follows the cited source as closely as possible. \label{tbl:SimTargetEsts}}
\end{table}

\begin{center}
\begin{tabular}{r|lcp{80mm}|c|}
No. & Type & Assumption & Model & Estimator \\
\hline
 & \multicolumn{4}{c|}{Calendar Period 3 Estimand} \\
\hline
1 & GD & S4 & Independent Covariance & $\hat{\theta}^{(4)}_3$ \\
2 & ME & S4 & $Y_{ijk} = \mu + \alpha_i + b_j + \nu_{ij} + \theta_j X_{ij} + \epsilon_{ijk}$ & $\hat{\theta}_3$ \\
3 & GD & S2 & Independent Covariance & $\frac{1}{2} \sum_{a=1}^2 \hat{\theta}^{(2)}_{3,a}$ \\
4 & ME & S2 & $Y_{ijk} = \mu + \alpha_i + b_j + \nu_{ij} + \theta_{j,a} X_{ij} I(T_i - j + 1 = a) + \epsilon_{ijk}$ & $\frac{1}{2} \sum_{a=1}^2 \hat{\theta}_{3,a}$ \\
\hline
 & \multicolumn{4}{c|}{Exposure Period 2 Estimand} \\
 \hline
5 & GD & S3 & Independent Covariance & $\hat{\theta}^{(3)}_2$ \\
6 & ME & S3 & $Y_{ijk} = \mu + \alpha_i + b_j + \nu_{ij} + \theta_a X_{ij} I(T_i - j + 1 = a) + \epsilon_{ijk}$ & $\hat{\theta}_2$ \\
7 & GD & S2 & Independent Covariance & $\frac{1}{5} \sum_{j=3}^7 \hat{\theta}^{(2)}_{j,2}$ \\
8 & ME & S2 & $Y_{ijk} = \mu + \alpha_i + b_j + \nu_{ij} + \theta_{j,a} X_{ij} I(T_i - j + 1 = a) + \epsilon_{ijk}$ & $\frac{1}{6} \sum_{j=3}^8 \hat{\theta}_{j,2}$ \\
\hline
\end{tabular}

\begin{tabular}{r|lcp{80mm}|c|}
No. & Type & Assumption & Model & Estimator \\
\hline
 & \multicolumn{4}{c|}{Exposure Period 1 Estimand} \\
 \hline
9 & GD & S3 & Independent Covariance & $\hat{\theta}^{(3)}_1$ \\
10 & ME & S3 & $Y_{ijk} = \mu + \alpha_i + b_j + \nu_{ij} + \theta_a X_{ij} I(T_i - j + 1 = a) + \epsilon_{ijk}$ & $\hat{\theta}_1$ \\
11 & SA & S2 & CO-2: proportional weights \newline (\cite{kennedyshaffer_novel_2020}) & $\hat{\theta}^{CO-2}$ \\
12 & GD & S2 & Independent Covariance & $\frac{1}{6} \sum_{j=2}^7 \hat{\theta}^{(2)}_{j,1}$ \\
13 & ME & S2 & $Y_{ijk} = \mu + \alpha_i + b_j + \nu_{ij} + \theta_{j,a} X_{ij} I(T_i - j + 1 = a) + \epsilon_{ijk}$ & $\frac{1}{6} \sum_{j=2}^8 \hat{\theta}_{j,1}$ \\
14 & SA & S2 & CH: first-difference estimator \newline (\cite{de_chaisemartin_two-way_2020}) & $\hat{DID}_M$ \\
15 & SA & S2 & CO-1: equal weights \newline (\cite{kennedyshaffer_novel_2020}) & $\hat{\theta}^{CO-1}$ \\
\hline
\end{tabular}
\end{center}

\clearpage

\section{Supplementary Results} \label{appx:extrares}

\begin{table}[!ht]
\caption{Odds ratio estimates and permutation test p-values for generalized DID estimators of summary effects of XpertMTB/RIF testing on tuberculosis outcomes in Brazil, 2012, under different variance assumptions. \label{tbl:Sres1}}
\begin{center}
\begin{tabular}{ll|r|r|r|r|r|r|}
& & \multicolumn{2}{c|}{} & \multicolumn{2}{c|}{Exchangeable,} & \multicolumn{2}{c|}{AR(1),} \\
& & \multicolumn{2}{c|}{Independence} & \multicolumn{2}{c|}{$\rho = 0.003$} & \multicolumn{2}{c|}{$\rho = 0.012$} \\
Assumption & Estimator & Estimate & P-Value & Estimate & P-Value  & Estimate & P-Value \\ \hline
S5 & $\hat{\theta}^{(5)}$ & 0.771 & 0.008 & 0.771 & 0.013 & 0.772 & 0.007 \\
S4 & $\frac{1}{6} \sum_{j=2}^7 \hat{\theta}^{(4)}_j$ & 0.784 & 0.026 & 0.784 & 0.021 & 0.785 & 0.029 \\
S3 & $\frac{1}{7} \sum_{a=1}^7 \hat{\theta}^{(3)}_a$ & 0.828 & 0.371 & 0.828 & 0.413 & 0.827 & 0.386 \\
S2 & $\frac{1}{21} \sum_{j=2}^7 \sum_{a=1}^{j-1} \hat{\theta}^{(2)}_{j,a}$ & 0.801 & 0.155 & 0.801 &  0.149 & 0.802 & 0.162 \\
\end{tabular}
\end{center}
\end{table}

\begin{table}
\caption{Risk difference estimates and permutation test p-values for generalized DID estimators of summary effects of XpertMTB/RIF testing on tuberculosis outcomes in Brazil, 2012, under different variance assumptions. \label{tbl:Sres2}}
\begin{center}
\begin{tabular}{ll|r|r|r|r|r|r|}
& & \multicolumn{2}{c|}{} & \multicolumn{2}{c|}{Exchangeable,} & \multicolumn{2}{c|}{AR(1),} \\
& & \multicolumn{2}{c|}{Independence} & \multicolumn{2}{c|}{$\rho = 0.003$} & \multicolumn{2}{c|}{$\rho = 0.012$} \\
Assumption & Estimator & Estimate & P-Value & Estimate & P-Value  & Estimate & P-Value \\ \hline
S5 & $\hat{\theta}^{(5)}$ & -7.35\% & 0.013 & -7.35\% & 0.020 & -7.33\% & 0.0013 \\
S4 & $\frac{1}{6} \sum_{j=2}^7 \hat{\theta}^{(4)}_j$ & -6.81\% & 0.054 & -6.81\% & 0.054 & -6.78\% & 0.048 \\
S3 & $\frac{1}{7} \sum_{a=1}^7 \hat{\theta}^{(3)}_a$ & -5.57\% & 0.409 & -5.57\% & 0.456 & -5.58\% & 0.430 \\
S2 & $\frac{1}{21} \sum_{j=2}^7 \sum_{a=1}^{j-1} \hat{\theta}^{(2)}_{j,a}$ & -6.11\% & 0.220 & -6.11\% &  0.229 & -6.09\% & 0.232 \\
\end{tabular}
\end{center}
\end{table}

\clearpage

\begin{figure}[!ht]
\begin{center}
\includegraphics[width=5in]{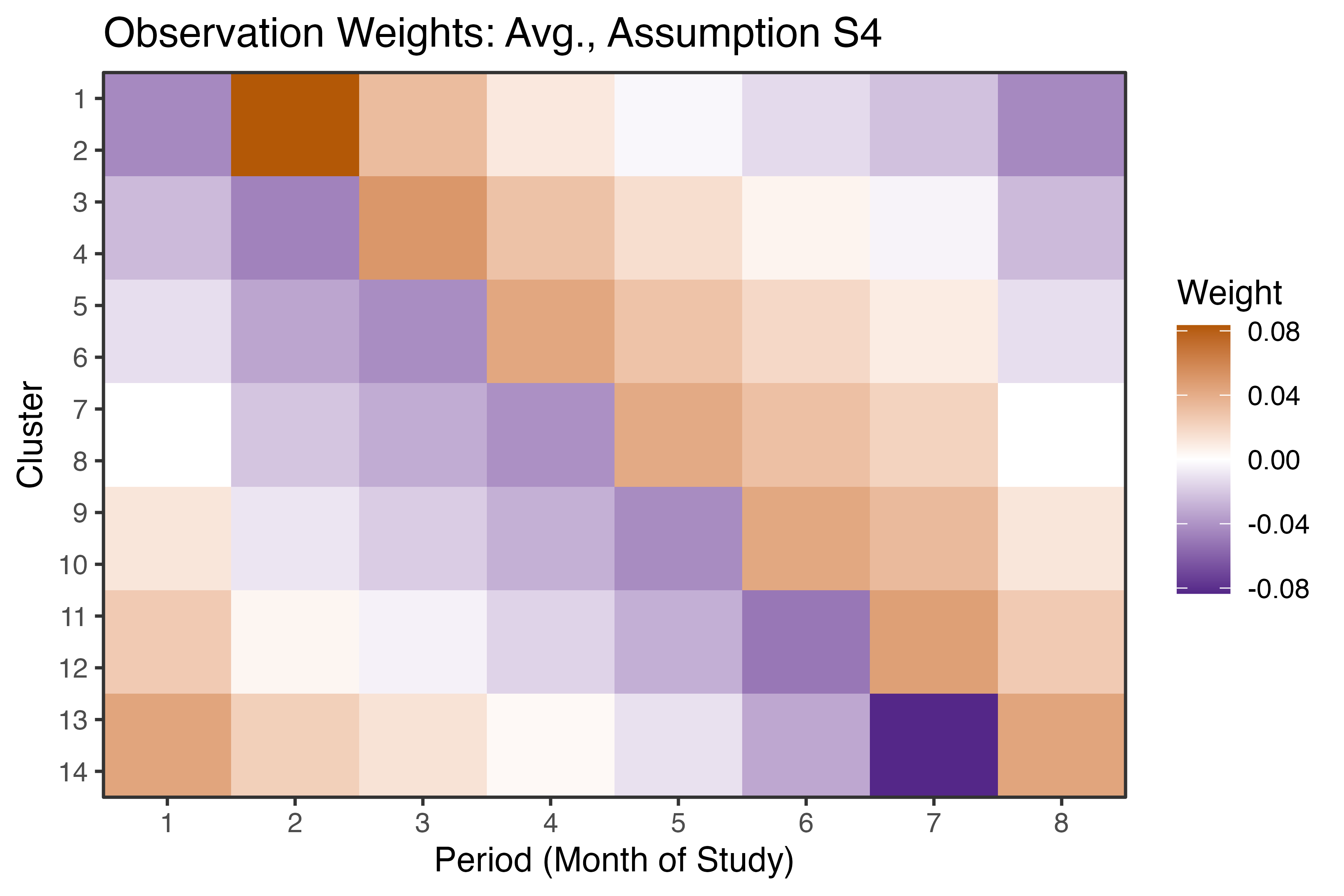}
\end{center}
\caption{Weights on cluster-period observations for the estimator $\frac{1}{6} \sum_{j=2}^7 \hat{\theta}^{(4)}_j$ fitted under an exchangeable variance structure with $\rho = 0.003$. \label{fig:obsWS4}}
\end{figure}

\clearpage

\begin{table}[!ht]
\caption{Cumulative percentage difference estimates and permutation test p-values for generalized DID estimators of summary effects of vaccine lotteries on COVID-19 vaccine uptake, United States Midwest region, May--July 2021. Estimation occurred using an independent working correlation structure with parameter $\rho = 0.95$.\label{tbl:vaxS1}}
\begin{center}
\begin{tabular}{ll|r|r|}
Description & Estimator & Estimate & P-Value  \\ \hline
Overall ATT & $\frac{1}{26} \sum_{\{j,a\}} \hat{\theta}_{j,a}^{(2)}$ & 1.318 & 0.265 \\
First-Period Effect & $\frac{1}{4} \sum_{j} \hat{\theta}_{j,1}^{(2)}$ & 1.311 & 0.044 \\
Second-Period Effect & $\frac{1}{4} \sum_{j} \hat{\theta}_{j,2}^{(2)}$ & 1.570 & 0.027 \\
Average Four-Week Effect & $\frac{1}{12} \sum_{\{j,a:~j-a \le 26, 1 \le a \le 4\}} \hat{\theta}_{j,a}^{(2)}$ & 1.424 & 0.136 \\
Average 2--4-Week Effect & $\frac{1}{9} \sum_{\{j,a:~j-a \le 26, 2 \le a \le 4\}} \hat{\theta}_{j,a}^{(2)}$ & 1.477 & 0.143 \\
State-Averaged Effect & $\frac{1}{4} \sum_{t=15}^{30} \frac{1}{31-t} \sum_{a=1}^{31-t} \hat{\theta}_{a+t,a}^{(2)}$ & 1.593 & 0.073 \\
Ohio Effect & $\frac{1}{12} \sum_{a=1}^{12} \hat{\theta}_{a+18,a}^{(2)}$ & -0.016 & 0.987 \\
Illinois Effect & $\frac{1}{7} \sum_{a=1}^{7} \hat{\theta}_{a+23,a}^{(2)}$ & 4.010 & 0.027 \\
\end{tabular}
\end{center}
\end{table}

\begin{figure}[!ht]
\begin{center}
\includegraphics[width=5in]{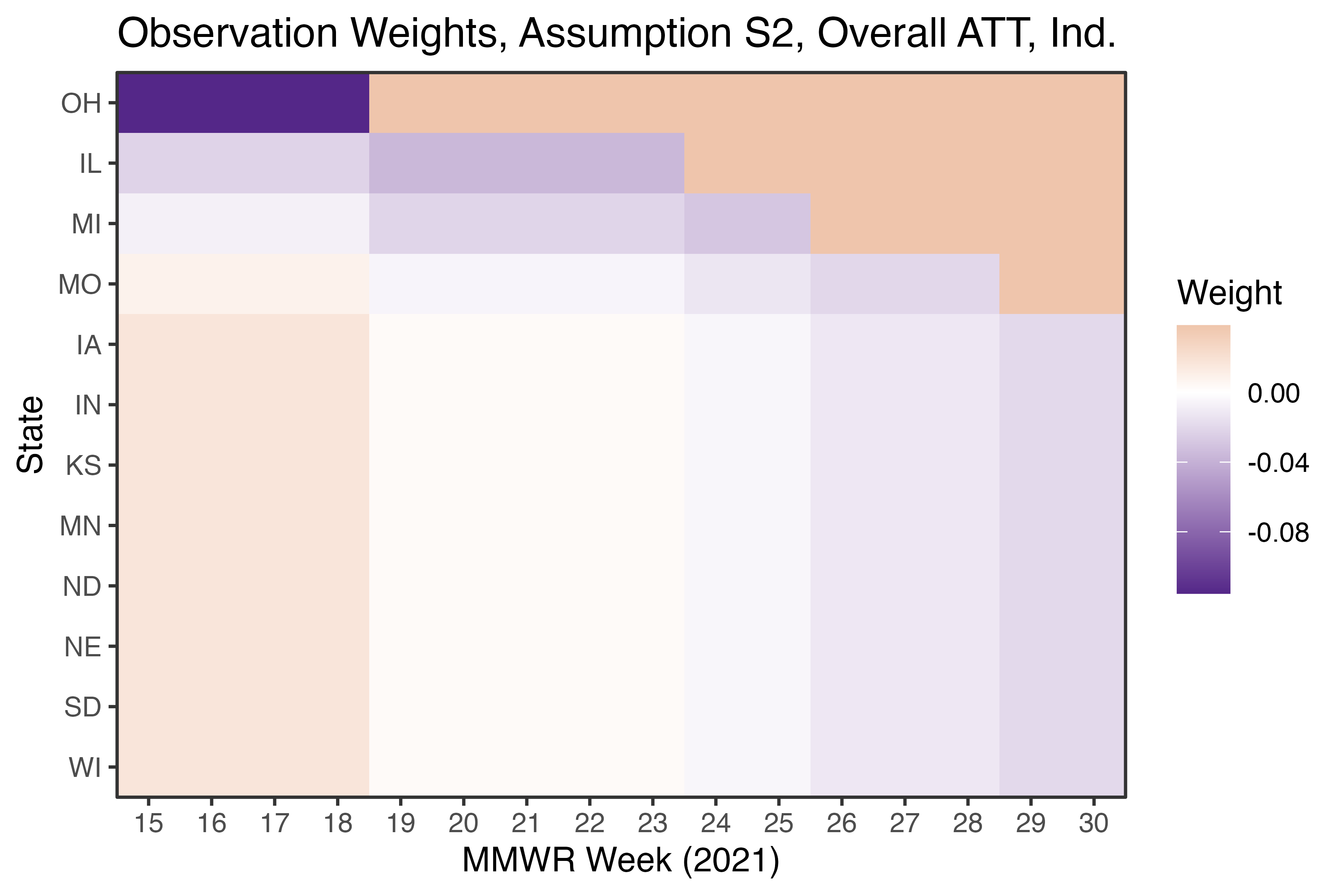}

\includegraphics[width=5in]{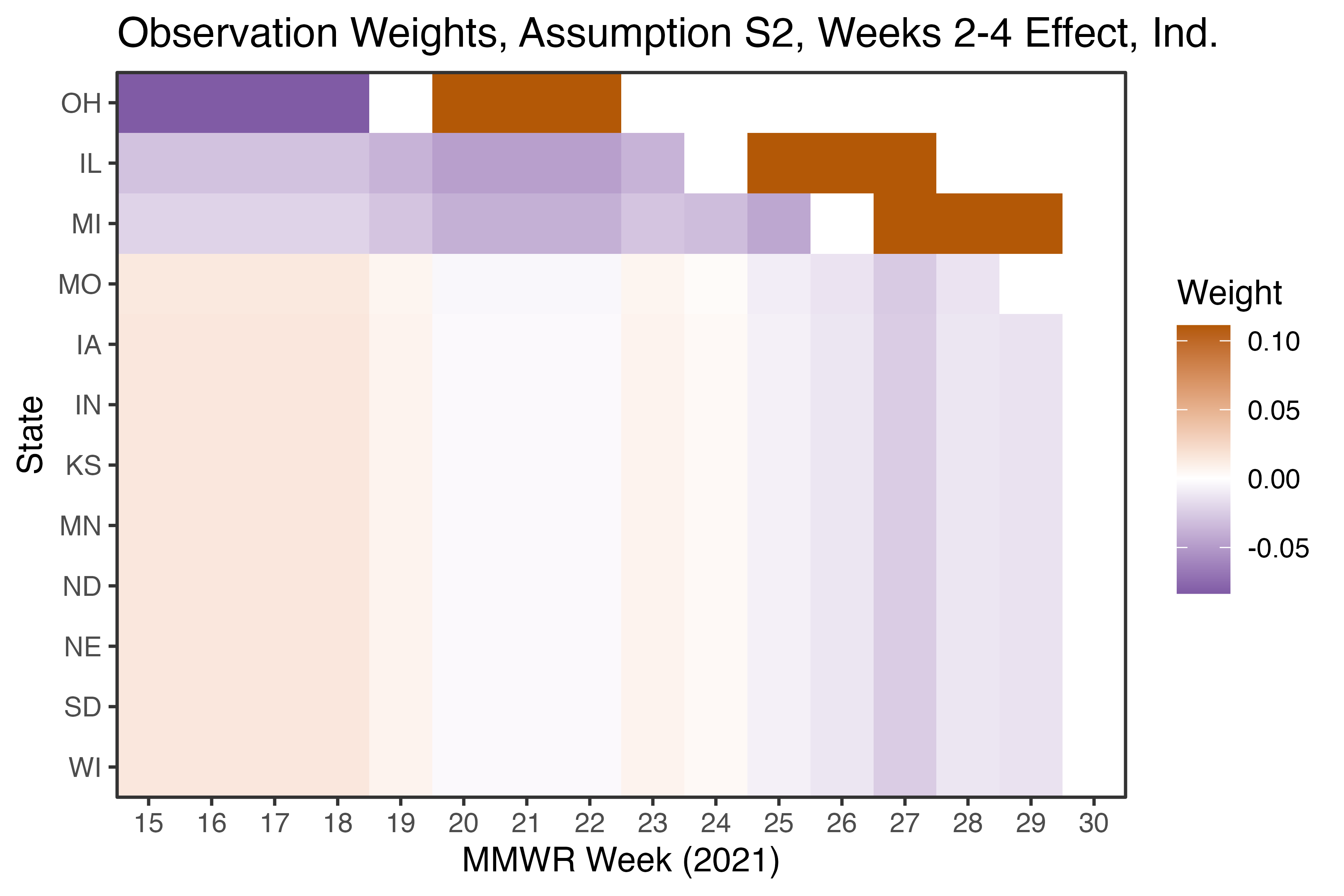}
\end{center}
\caption{Weights on cluster-period observations for the overall ATT estimator (top) and average 2--4-week effect estimator (bottom) under assumption S2 fitted under an independent working correlation structure. \label{fig:vaxWS1}}
\end{figure}

\clearpage

\begin{figure}[!ht]
\begin{center}
\includegraphics[width=7in]{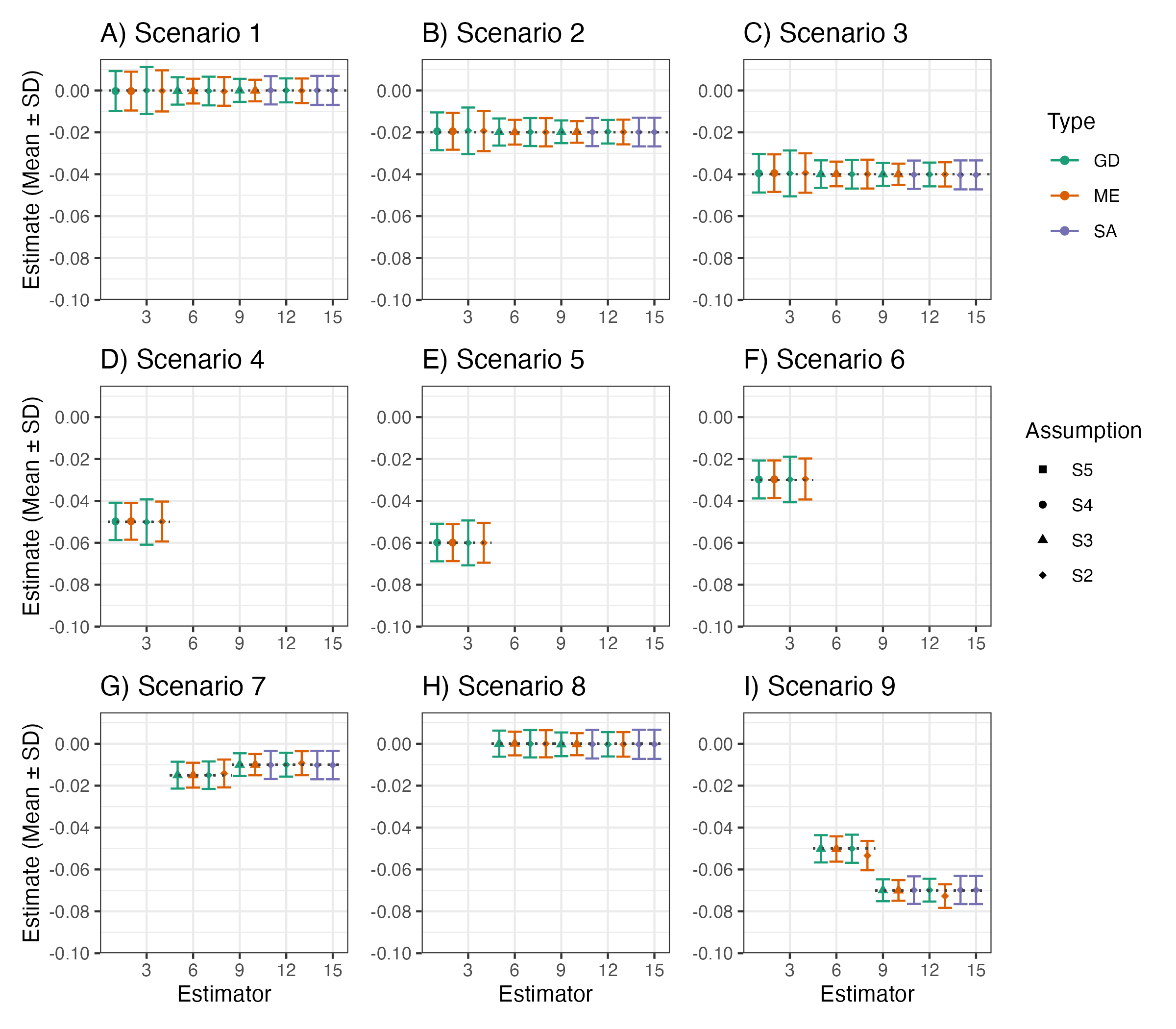}
\end{center}
\caption{Simulation results (mean estimate plus or minus one standard deviation of estimates across simulations) for 1,000 simulations each, by simulation scenario and estimator for period-specific estimators. Dotted horizontal lines represent the true value of the target estimand.\label{fig:sim_t_ests}}
\end{figure}

\begin{figure}[!ht]
\begin{center}
\includegraphics[width=7in]{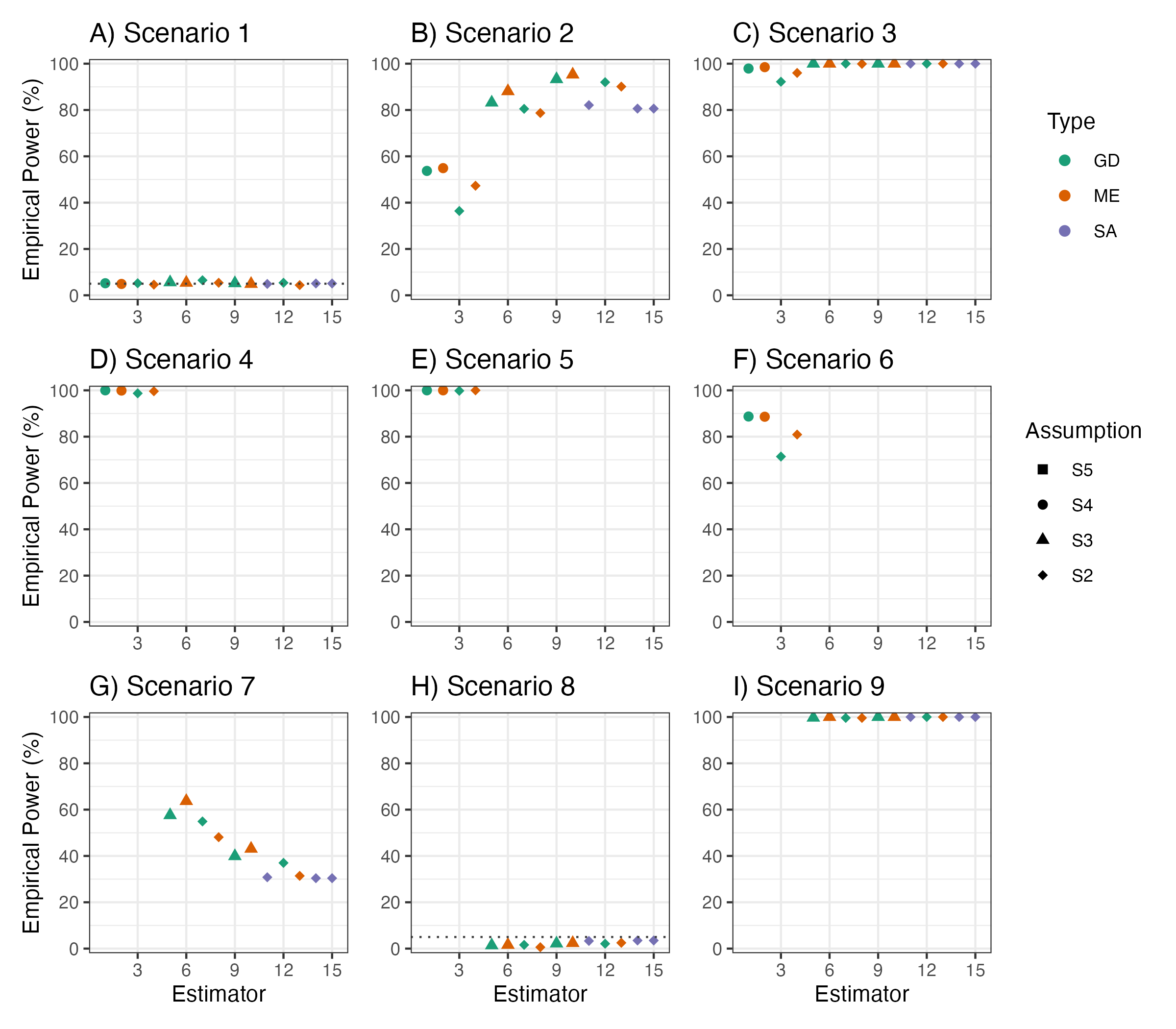}
\end{center}
\caption{Simulation results (empirical power at $\alpha=0.05$ level) for 1,000 simulations each, by simulation scenario and estimator for period-specific estimators. Note that biased estimators or those targeting substantially different estimands are excluded.\label{fig:sim_t_power}}
\end{figure}


%
%
%
%
\end{document}